\documentclass[a4paper,11pt]{article}
\usepackage{jinstpub} 
\usepackage{subcaption}%
\usepackage{multirow}%
\usepackage{float}%
\usepackage{orcidlink}%
\usepackage{lineno}

\title{\boldmath Development of A Novel Compton Camera for MeV Gamma-Ray Measurement in Space}

\author[a]{Pingwei Sun\orcidlink{0009-0003-6749-9360},}
\author[a]{Wenxiang Fang,}
\author[a]{Guorong He\orcidlink{0009-0008-4353-1019},}
\author[a]{Zhen Wu\orcidlink{0009-0008-1643-4002},}
\author[a]{Jinghe Yang\orcidlink{0009-0005-3134-5225},}
\author[b, a]{Jiancheng Zeng\orcidlink{0000-0002-4157-1964},}
\author[a]{Yiyu Pan\orcidlink{0009-0005-3205-2221},}
\author[a]{Jiacheng Ding,}
\author[c]{Enzhao Qi\orcidlink{0009-0005-2993-2850},}
\author[a]{Jiahao Su,}
\author[a]{Haoran Yang,}
\author[a]{Bowen Zhu,}
\author[a, 1]{Mengjiao Xiao\orcidlink{0000-0002-6397-617X},\note{Corresponding author.}}

\affiliation[a]{State Key Laboratory of Dark Matter Physics, Key Laboratory for Particle Astrophysics and Cosmology (MoE), Shanghai Key Laboratory for Particle Physics and Cosmology, School of Physics and Astronomy, Shanghai Jiao Tong University,\\800 Dongchuan Road, Shanghai 200240, China}
\affiliation[b]{State Key Laboratory of Dark Matter Physics, Key Laboratory for Particle Astrophysics and Cosmology (MoE), Shanghai Key Laboratory for Particle Physics and Cosmology, Tsung-Dao Lee Institute, Shanghai Jiao Tong University,\\1 Lisuo Road, Shanghai 201210, China}
\affiliation[c]{School of Ocean and Civil Engineering, Shanghai Jiao Tong University,\\800 Dongchuan Road, Shanghai 200240, China}

\emailAdd{xmjzcc@sjtu.edu.cn}

\abstract{The astrophysical gamma rays in the MeV energy region have not yet been well-explored due to the limitation of detection technology in the past decades, and the famous gamma-ray “MeV gap” exists. Opening the window of MeV gamma-ray is not only critical for the gamma astronomy but also essential for rich frontier researches in astro-particle physics, such as detecting light dark matter, probing the primordial black hole and better understanding of nucleosynthesis. As a pilot experiment of the project for dark matter detection in space at Shanghai Jiao Tong University, a three-layer Compton camera with the energy resolution better than 4\% and position resolution of $\sim$1 mm is developed utilizing the novel scintillators. Here we show the design, detailed calibration and validation results of the novel Compton camera, and demonstrate its good ability of MeV gamma-ray source imaging for the upcoming in-orbit mission.}

\keywords{Gamma detectors (scintillators, CZT, HPGe, HgI etc); Compton imaging; Dark Matter detectors (WIMPs, axions, etc.)} 

\arxivnumber{2608.21216} 

\begin{document}
\maketitle
\flushbottom

\section{Introduction}
Photons are the most abundant particles in the Universe and play a crucial role in researches of the matter world. Through observations of photons at different energies, people have achieved remarkable accomplishments in cosmology, astrophysics and high-energy particle physics. In gamma-ray astronomy, MeV gamma-rays are key to understanding the transition of the gamma-ray spectrum from low energies (keV) to high energies (GeV). A precise measurement of MeV gamma-rays provides an unique window to study dark matter annihilation \cite{ODonnell2025MeVDM,Guo:2023kqt,dm2025}, Hawking radiation from primordial black holes \cite{Coogan2021PBH}, gamma-ray bursts \cite{Piran2005GRB}, Galactic nucleosynthesis and chemical evolution \cite{Diehl2006Al26}, etc. In addition, MeV gamma-rays could be produced in association with gravitational waves from neutron star mergers or with neutrinos from primordial black holes, providing an important means for multi-messenger astronomy \cite{Abbott2017MultiMessenger}.

However, the photons at the MeV energy range interact with matter mainly via the Compton scattering with low cross sections which makes its measurement challenging. The current observation sensitivity of astrophysical photons in the MeV band is significantly lower than that in X-rays and high-energy GeV gamma-rays, resulting in the famous "MeV Gap" problem \cite{W.Atwood2009,F.A.Harrison2013,Knodlseder2016GammaFuture,DeAngelis2017eASTROGAM,Ackermann2025MeV}. A Compton camera is a gamma-ray detector which utilizes the kinematics of Compton scattering to image a gamma trajectory. The concept was initially proposed around 50 years ago \cite{SCHONFELDER1973385} and later on widely used in high-energy astrophysics, in particular for the MeV gamma-ray observation in space. COMPTEL (The Imaging COMPton TELescope) on the Compton Gamma-Ray Observatory (CGRO) ‌pioneeringly observed around 30 sources at MeV energy \cite{V.Schönfelder2000}, while the INTEGRAL \cite{Winkler2003INTEGRAL} as the CGRO's successor is in-orbit operation, but the sensitivity is still limited. Recently, many next-generation satellite experiments specifically targeting MeV gamma-rays have been proposed. These planned experiments use different kinds of detectors but the basic detection principle is using the technology of Compton scattering and e$^{+}$e$^{-}$ pair-production for MeV gamma observation, such as the COSI experiment \cite{Beechert2022COSI}, AMEGO/AMEGO-2 \cite{Caputo2022AMEGOX}, e-ASTROGAM \cite{DeAngelis2017eASTROGAM, DeAngelis2021ASTROGAM}, GRID-MASS \cite{Wen2019GRID}, MeGaT \cite{zhao2025megat}. In addition to these proposed satellite missions, the balloon experiments such as GRAMS \cite{Aramaki2020GRAMS} and SMILE-3 \cite{Takada2022SMILE2plus} also focus on gamma-ray measurement in this energy range. Besides those specific experiments, the VLAST \cite{Pan2024VLAST} and HERD \cite{Kyratzis2023HERD} experiment primarily aim at high energies but could potentially probe the gamma-rays down to hundreds of MeV. These next-generation experiments with higher sensitivities are under development and some of them are expected to be commissioning in the coming years. As part of the global efforts to push a deeper detection of the MeV gamma-rays in space, we propose to develop a highly compact and modular Compton camera using novel scintillators with a competitive projected sensitivity in the energy range of 0.2$\sim$3 MeV. This novel Compton camera keeps the capability of upgrade by modularization and thus serves as a pathfinder of the dark matter search mission in space at Shanghai Jiao Tong University which is strongly complementary to the existing and planned projects for the precise measurement of MeV gamma-rays in future.

This work reports on the R$\&$D efforts to develop a three-layer Compton camera for observing the MeV gamma-rays in space. The Compton camera design and construction are briefly described in Section~\ref{sec:design}. The calibration strategy and results are detailed in Section~\ref{sec:calibration}. In Section~\ref{sec:performance}, we discuss the instrumental performances at the system level including the track and angular reconstruction for gamma imaging. The critical environmental tests for the instrument operation in space are reported in Section~\ref{sec:validation}. Finally, conclusions and future prospects are presented in Section~\ref{sec:conclusion}.

\section{Compton Camera Design \& Construction}
\label{sec:design}
\begin{figure}[htbp]
\centering
    \begin{subfigure}[b]{0.45\textwidth}
        \includegraphics[width=\linewidth]{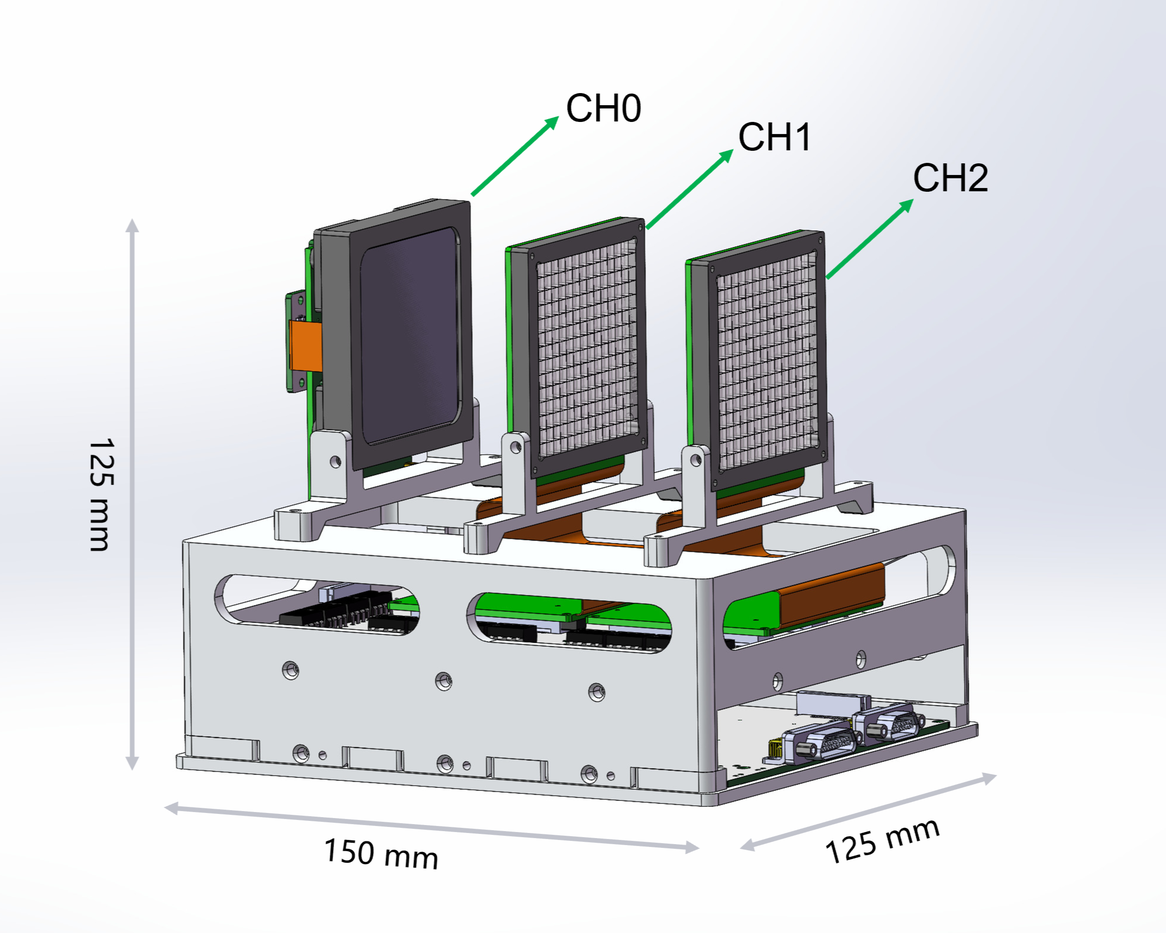}
        \caption{Design}
        \label{fig:camera_design}
    \end{subfigure}
    \quad
    \begin{subfigure}[b]{0.45\textwidth}
        \includegraphics[width=\linewidth]{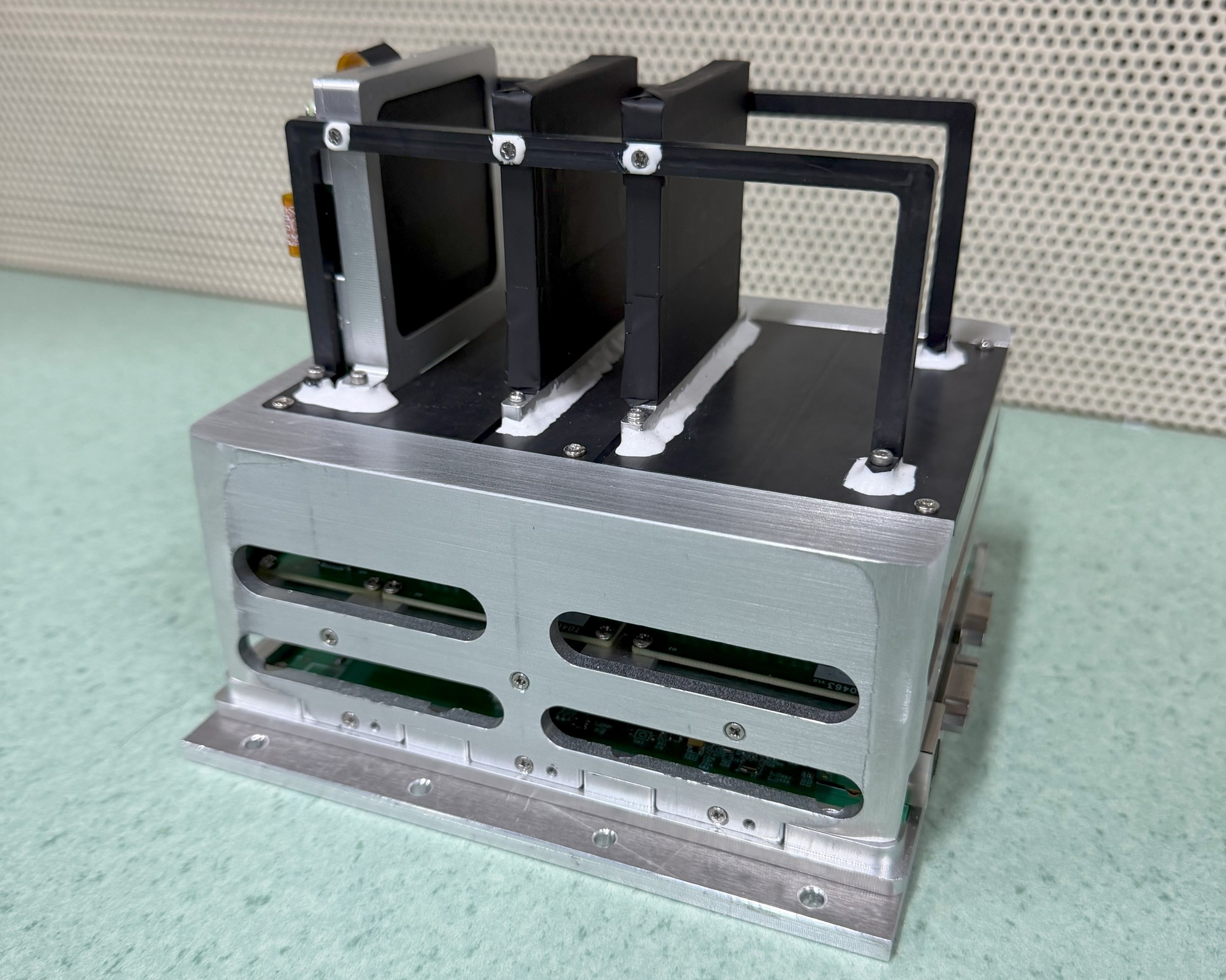}
        \caption{Photograph}
        \label{fig:camera_real}
    \end{subfigure}
\caption{Compton camera: design and photograph (the aluminum cover which encloses the detectors is not shown).}
\label{fig:camera}
\end{figure}

The Compton camera developed in this work targets to measure the gamma-rays in the energy range of 0.2$\sim$3 MeV with a field-of-view larger than 45$^{\circ}$. In order to achieve a sensitivity competitive with the existing experiments, the energy resolution of the Compton camera is required to be better than 5\% at 1 MeV, the position resolution better than 3 mm and angular resolution better than 10$^{\circ}$. In addition, this Compton camera is designed to share a commercial microsatellite with other payloads, so its overall dimensions are limited by 150$\times$150$\times$150 mm and the total power consumption is required to be less than 10 W. The design of the Compton camera is shown in Figure~\ref{fig:camera_design}. It is highly compact with the overall dimension of $\sim$150$\times$125$\times$125 mm (L$\times$W$\times$H) and total weight of $\sim$1.75 kg. The detectors are mounted on top while the electronics are underneath, both are mechanically fixed on the aluminum frame with stainless steel screws. In addition, all the components are enclosed in an aluminum cover (which is not shown in Figure~\ref{fig:camera_design}). We detail the mechanical design and the detector electronics in this section.

\subsection{Mechanical Design}
\label{subsec:structure}
We choose the novel scintillation materials to build the detectors of Compton camera. The selected scintillators have high resistance to the radiation. In addition, they do not deliquesce which helps the detector robust against the environmental conditions during the ground testing, and essentially ensures the detector's performances consistent in the ground calibration and in-orbit operation. These features make them the ideal materials for detectors operated in space. The Compton camera developed here employs three detector layers labeled as CH0, CH1 and CH2 respectively, as illustrated in Figure~\ref{fig:camera_design}. CH1 and CH2 are the scattering layers and CH0 serves as an absorption layer. All the three layers have the dimensions of 47.04$\times$47.04 mm while the scattering layers are 3 mm in thickness but the absorption layer is 6 mm. 

By design, the two scattering layers are identical with each other. The scattering layer uses Cerium doped Silicate Yttrium (Y$_{2}$SiO$_{5}$:Ce, hereafter simply YSO) scintillation crystals. YSO has a moderate density ($\rho$$\approx$4.5 g/cm$^{3}$) and features a high light yield ($\sim$28000 photons/MeV) with a wavelength peaked at $\sim$410 nm and short decay time~\cite{Chewpraditkul2012YSOLYSO, YSOdecaytime, yoshikawa2013czochralski}, enabling a rapid response to the scattered events from the incident gamma-ray signals. Each scattering layer comprises 196 YSO cells which are uniformly integrated into an array of 14$\times$14 pixels. All the YSO pixels used in the scattering layers were fabricated from the same crystal ingot in order to assure the detector's homogeneity. Five sides of the crystal cell are coated with the BaSO$_{4}$ reflection layer to increase the light collection efficiency and the opening side of the crystal cell is coupled with a 3.36$\times$3.36 mm Silicon Photomultipliers (SiPMs).

The absorption layer utilizes Lutetium Yttrium Silicate:Cerium (hereafter simply LYSO) scintillation crystal. Compared to the YSO material used in the scattering layer, LYSO scintillator offers an even higher light yield ($\sim$30000 photons/MeV) with a wavelength peaked at $\sim$420 nm and larger radiation energy absorption capacity benefiting from its high density ($\rho$$\approx$7.2 g/cm$^{3}$)~\cite{Chewpraditkul2012YSOLYSO,Mao2014LYSO}. Its increased thickness and density make a high absorption efficiency for gamma-rays. The absorption layer consists of a single continuous crystal bulk. Its four sides are coupled with the strip array of 4$\times$14 SiPMs, and the other two sides are coated with the BaSO$_{4}$ foils as the photon reflectors.

The distances between the detector layers were optimized by simulations with the Geant4~\cite{Agostinelli2003Geant4}, by taking into account of the detection acceptance, angular resolution as well as the effect of intrinsic radioactivity from the CH0. Based on the requirements of field-of-view and angular resolution discussed previously, the distance between CH2 and CH1 is set at 25 mm. Furthermore, to suppress the background contribution from $^{176}$Lu to the level of 10$^{-3}$ events/s/keV while maintaining the required geometric acceptance, CH0 is then separated from CH1 by 35 mm. The electronic components (which will be discussed separately) are enclosed underneath the scintillators and a plastic board is placed in between as a thermal insulator. Figure~\ref{fig:camera_real} shows the Compton camera that has been constructed.

\subsection{Front-end Electronics \& DAQ}
\label{subsec:daq}
The deposited energy from the particle interaction inside the LYSO/YSO produces the scintillation photons. These photons are detected by the SiPMs. As described in the previous section, 448 SiPMs in total are used in this Compton camera. However, consider the energy range of interest and the size of scintillator cell, the produced photons are supposed to be very locally clustered. In this scenario, only small amount of SiPMs are expected to be fired at the same for an individual event. In addition, there is the severe requirement of power consumption for the detectors operated in space. Combining the above facts, a capacitive charge-division multiplexing circuit is adopted to read out the signals from the SiPMs~\cite{Choe2017}. 

Unlike the traditional resistive multiplexing networks which might degrade the timing performance due to low-pass filtering effects, the capacitive charge-division multiplexing scheme that we are using reduces the channel count through an optimized, weighted capacitor network, while maintaining high timing and energy resolution \cite{Downie2013Multiplexing}. Specifically, after the conversion through this network and the assignment of weights, the channels from a SiPM array for each detector layer are compressed into a four-channel readout system. These four positional signals, labeled as P, Q, R and S, are coded with both energy and position information \cite{Choe2017}.

A custom data acquisition (DAQ) system  based on the Field Programmable Gate Array (FPGA) is developed to process the signals from each detector layer. It employs an enhanced high-speed fully differential amplifier to process the raw analog signals and an enhanced low-noise analog-to-digital converter (ADC) with quad channels, 14 Bit and 125 MS/S (mega-samples per second) to convert the analog signals to digital values (ADC). To interface between the Compton camera and the satellite, a Controller Area Network (CAN) bus using a flexible-data-rate transceiver with shutdown and standby mode is utilized to receive telemetry and remote sensing commands while two mutually redundant Low-Voltage Differential Signaling (LVDS) lines are implemented to transmit the data. Those critical electronic components (FPGA, operational amplifiers, ADC chips, etc.) are selected to be suitable for the aerospace applications. They have been qualified by the previous satellite projects: radiation-tolerant, thermally resilient, mechanically robust and power-efficient. The mitigation to the single-event upset is implemented in the design for FPGA application in space. In addition, the space-grade printed circuit boards are fabricated to ensure a reliable in-orbit operation. The comprehensive environmental tests have been conducted to validate the selected components, see more detailed discussions in Section~\ref{sec:validation}.

‌In practice, the bias voltages for the SiPMs and shaping time for the signal process are nailed down by optimizing the energy resolution and range with the dedicated calibration. These parameters are then configured in the detectors via the CAN bus. During data processing, the amplified signal is split into two paths: one is for energy position sampling and the other for triggering. Through the FPGA, each detector is configured with its own gate discrimination which includes a low threshold to filter the noise and a high threshold to veto the high-energy events. The custom DAQ system contains two trigger modes: calibration mode and working mode. In the calibration mode, the DAQ collects the data if any detector has the hit in the required energy window. A coincidence trigger is designed for the working mode in which the DAQ acquires the data when CH1 and CH2 are fired simultaneously with the hit in the required energy. In both trigger modes, the hits for all three detectors are acquired and the data are saved in the format with an event ID header, UTC time stamp plus the internal time clock, P/Q/R/S value for CH0, CH1 and CH2, respectively.

For each individual event, the positional signals (PQRS) are used to reconstruct the XY position in the offline data analysis based on the following approaches~\cite{Choe2017}. Given that CH1/CH2 is equipped with an array of $14\times14$ SiPMs, a standard center-of-gravity algorithm is employed to reconstruct the interaction positions $(X_{i}, Y_{i})$, where $i \in \{1, 2\}$:
\begin{equation}
\label{eq:position12}
X_{i} = \frac{(R_i + S_i) - (P_i + Q_i)}{P_i + Q_i + R_i + S_i}, 
\qquad
Y_{i} = \frac{(Q_i + R_i) - (P_i + S_i)}{P_i + Q_i + R_i + S_i}
\end{equation}
for CH0, an array of 1$\times$14 SiPMs is mounted on each side of the crystal. Therefore, the \((X_{0},Y_{0})\) position is reconstructed by counting the differential amount of photons measured at the two opposite sides, as described below:
\begin{equation}
\label{eq:position0}
X_{0}=\frac{P_0 - R_0}{P_0 + R_0}, \qquad Y_{0}=\frac{Q_0 - S_0}{Q_0 + S_0}
\end{equation}

\section{Detector Calibration}
\label{sec:calibration}
The detector calibration includes the energy scale calibration which builds the detector's readouts from the electronics (in the unit of ADC) to the deposited energies (in the unit of keV), and the non-uniformity calibration which maps the detector's position-dependent response. 

\subsection{Calibration Sources \& Data}
\begin{figure}[htbp]   
    \centering        
    \begin{subfigure}[t]{0.25\textwidth}   
        \includegraphics[width=\linewidth]{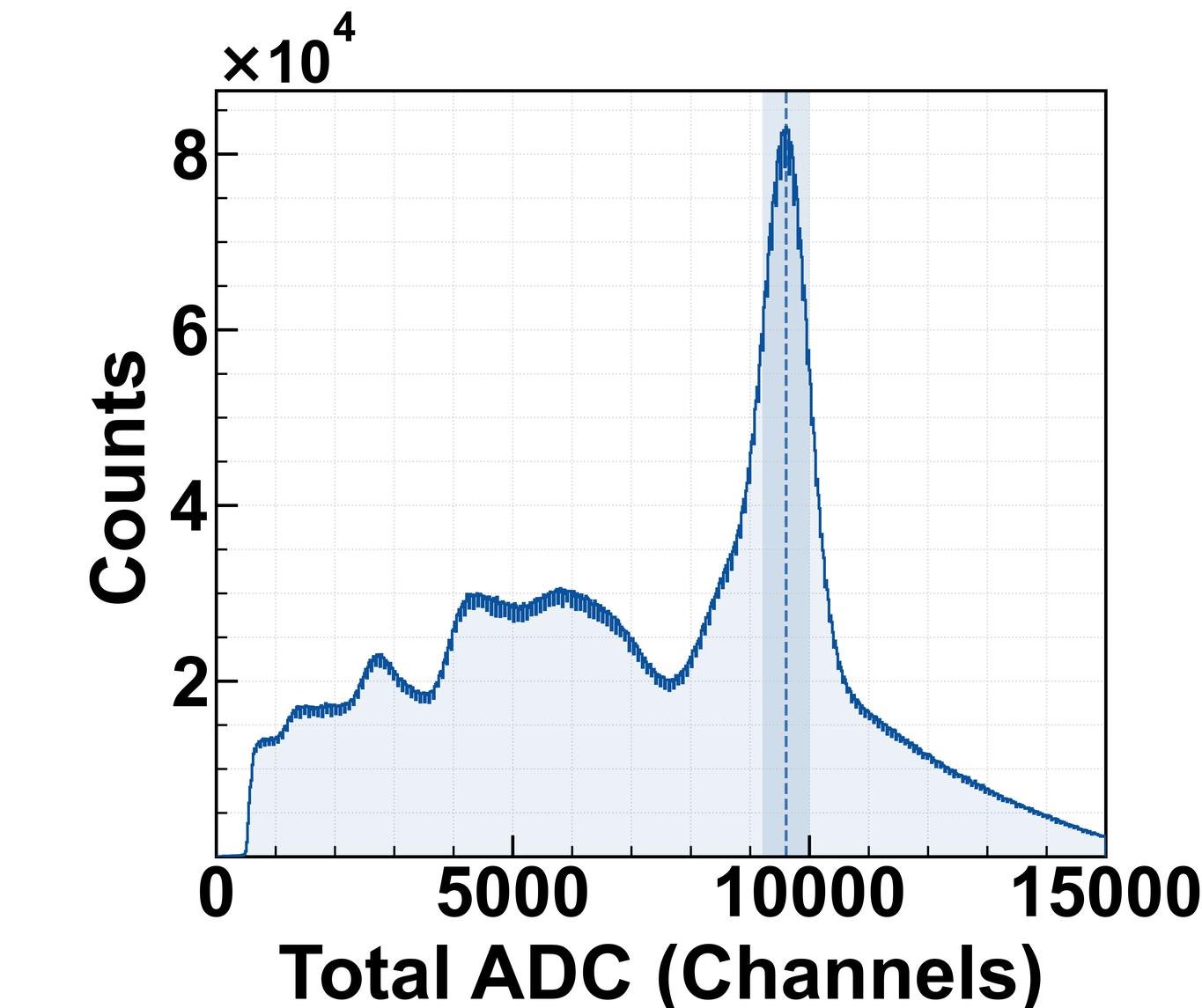}
        \caption{}
        \label{fig:radio_spec_cs_det0}
    \end{subfigure}
    \quad
    \begin{subfigure}[t]{0.25\textwidth}
        \includegraphics[width=\linewidth]{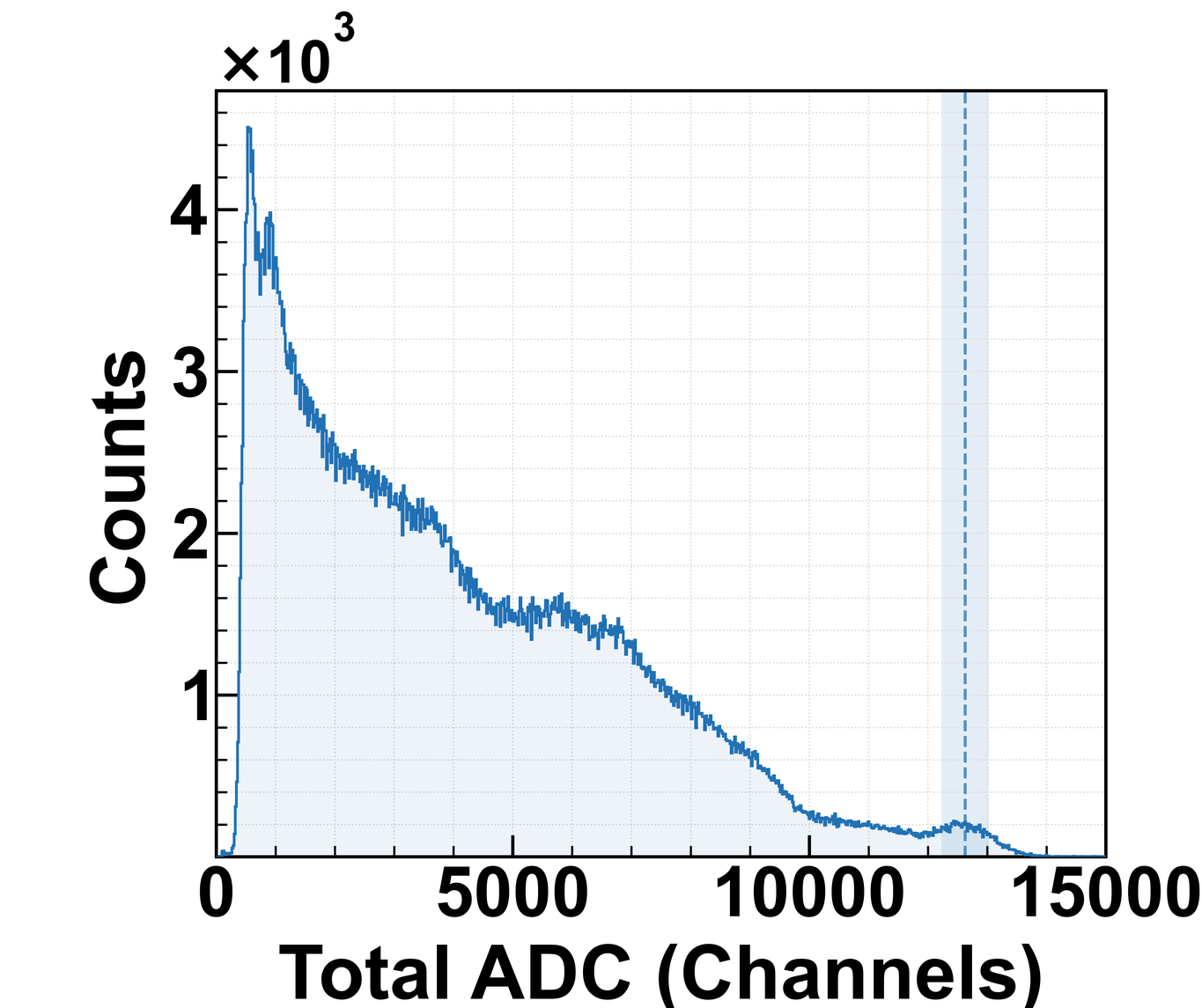}
        \caption{}
        \label{fig:radio_spec_cs_det1}
    \end{subfigure}
    \quad
    \begin{subfigure}[t]{0.25\textwidth}
        \includegraphics[width=\linewidth]{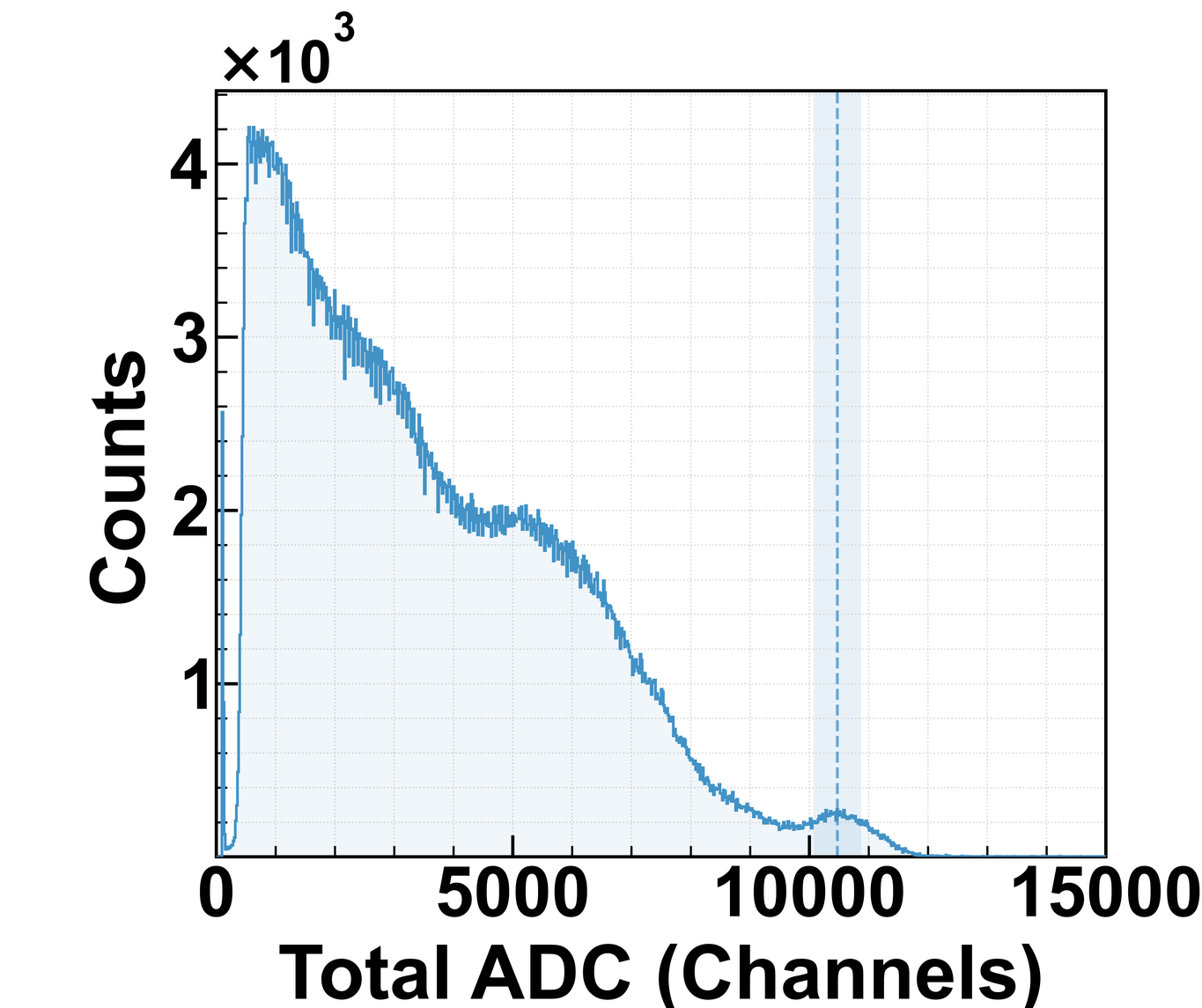}
        \caption{}
        \label{fig:radio_spec_cs_det2}
    \end{subfigure}  
    \par 
    \begin{subfigure}[t]{0.25\textwidth}
        \includegraphics[width=\linewidth]{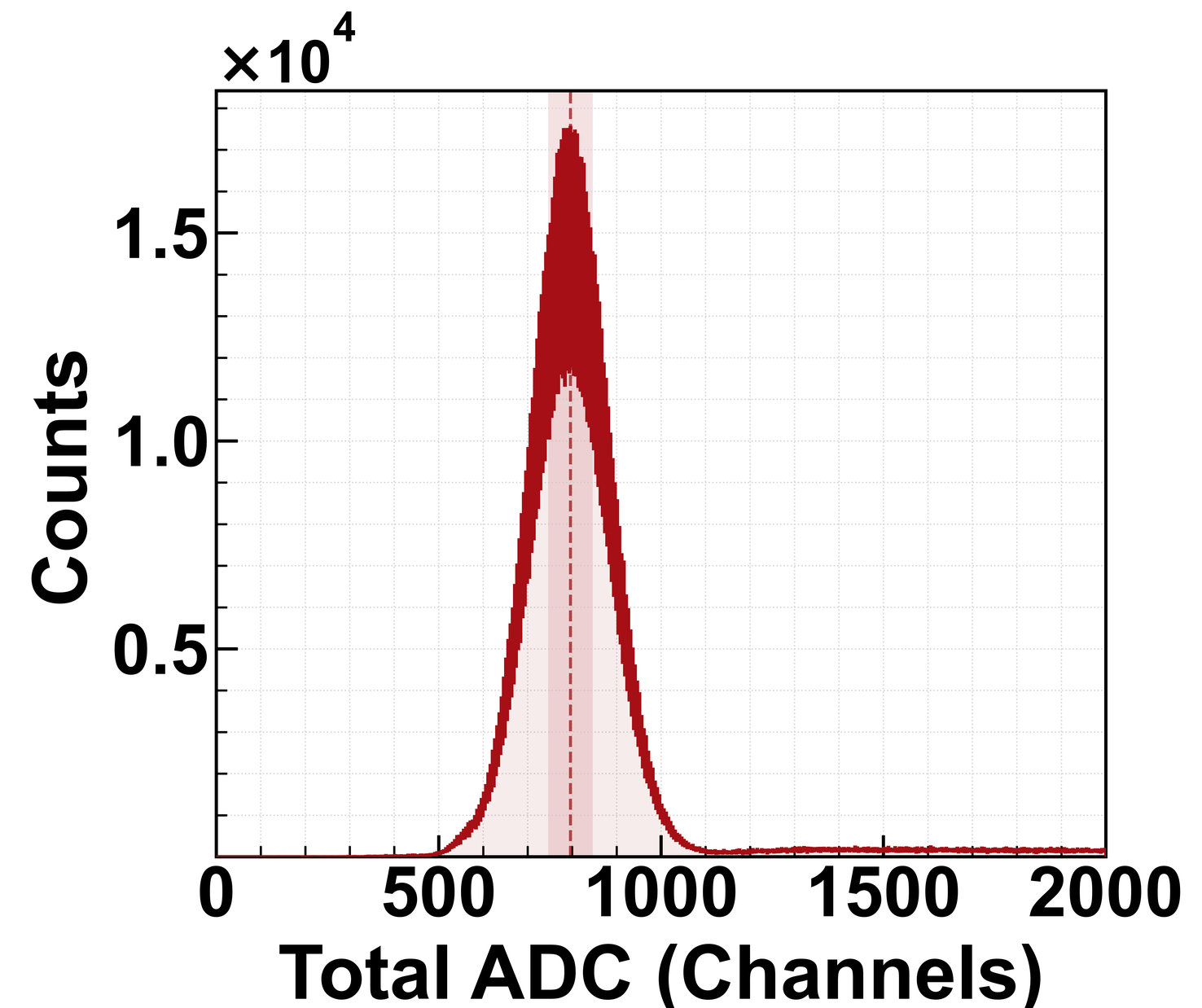}
        \caption{}
        \label{fig:radio_spec_am_det0}
    \end{subfigure}
    \quad
    \begin{subfigure}[t]{0.25\textwidth}
        \includegraphics[width=\linewidth]{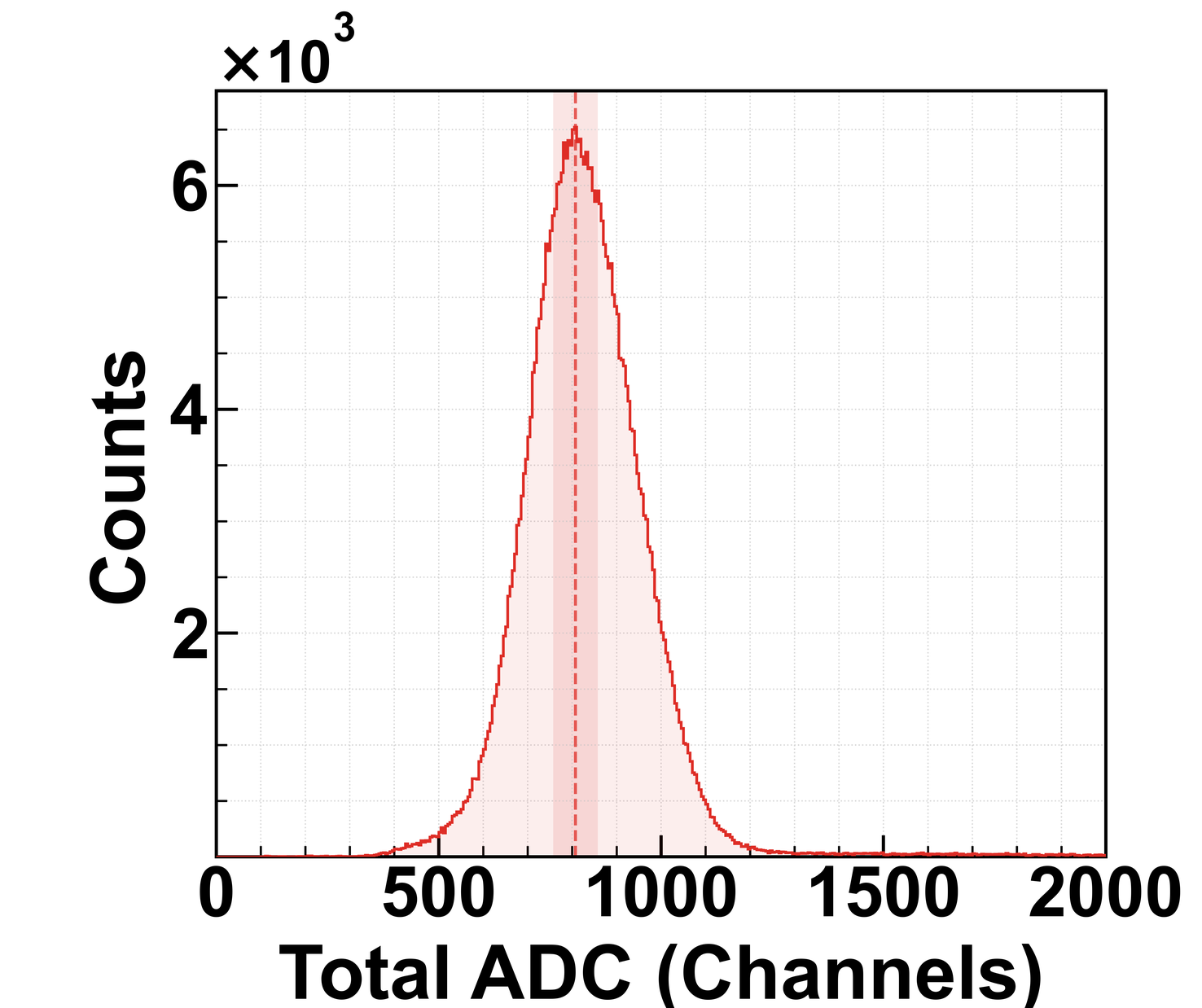}
        \caption{}
        \label{fig:radio_spec_am_det1}
    \end{subfigure}
    \quad
    \begin{subfigure}[t]{0.25\textwidth}
        \includegraphics[width=\linewidth]{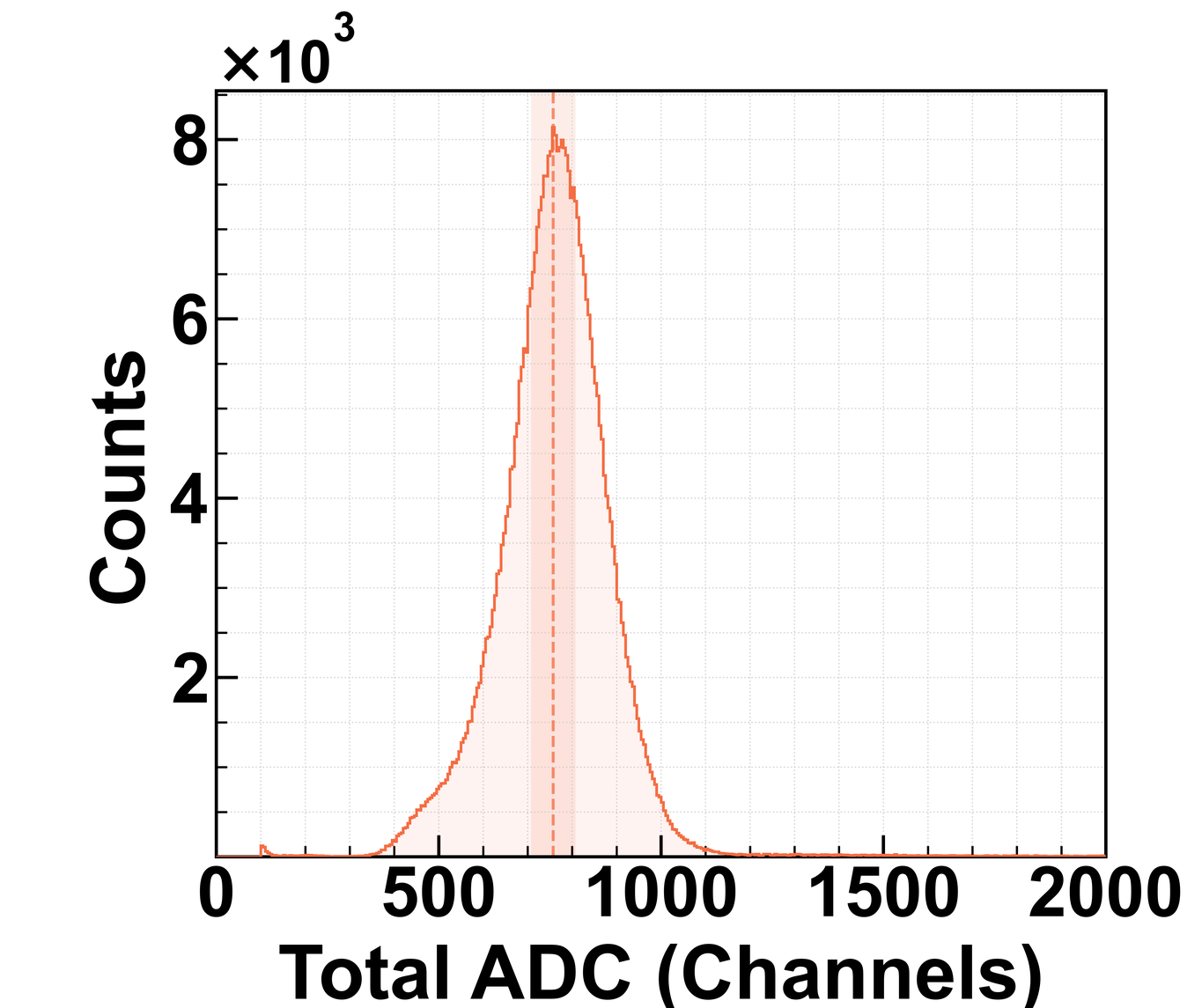}
        \caption{}
        \label{fig:radio_spec_am_det2}
    \end{subfigure}
    \par 
    \begin{subfigure}[t]{0.25\textwidth}
        \includegraphics[width=\linewidth]{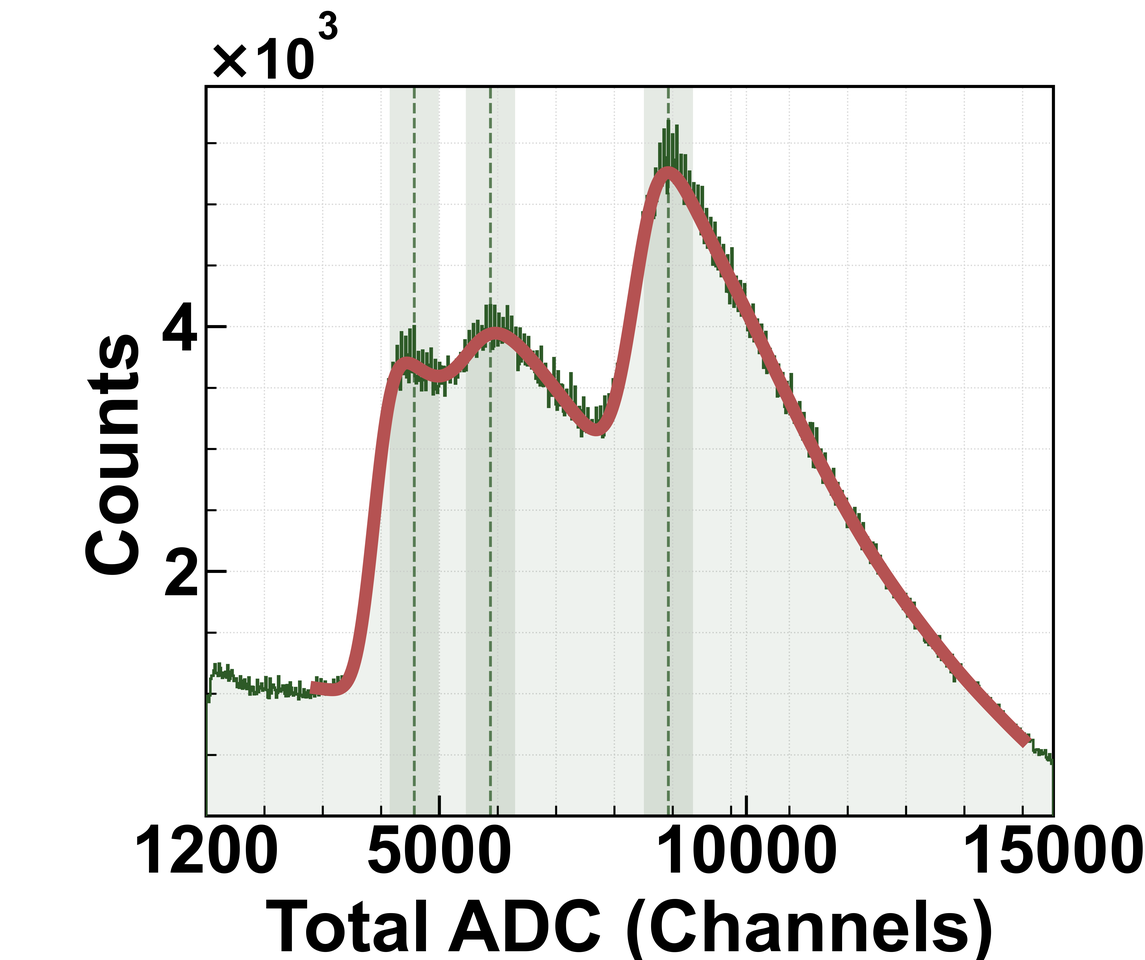}
        \caption{}
        \label{fig:radio_spec_lu_det0}
    \end{subfigure}
    \quad
    \begin{subfigure}[t]{0.25\textwidth}
        \includegraphics[width=\linewidth]{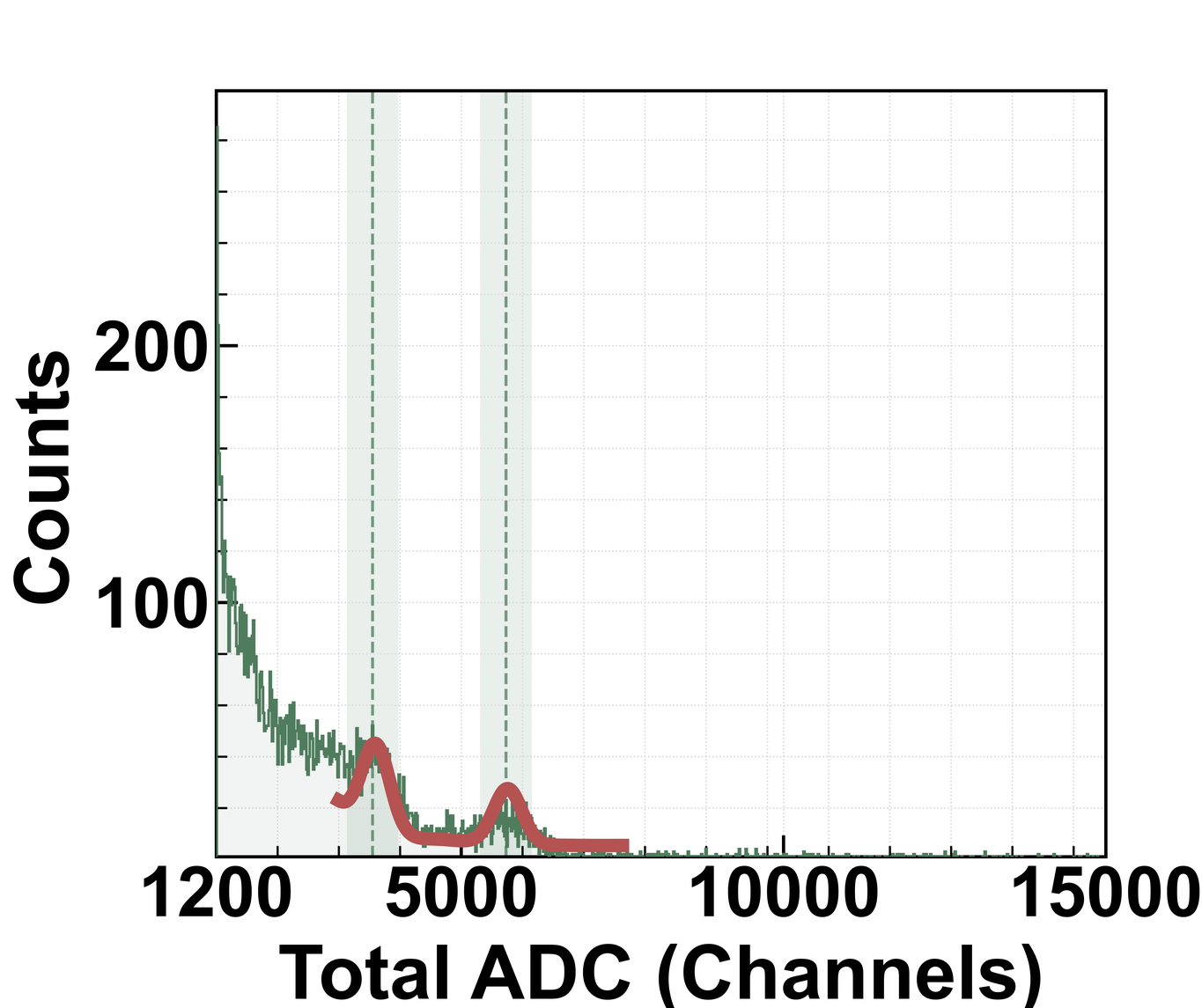}
        \caption{}
        \label{fig:radio_spec_lu_det1}
    \end{subfigure}
    \quad
    \begin{subfigure}[t]{0.25\textwidth}
        \includegraphics[width=\linewidth]{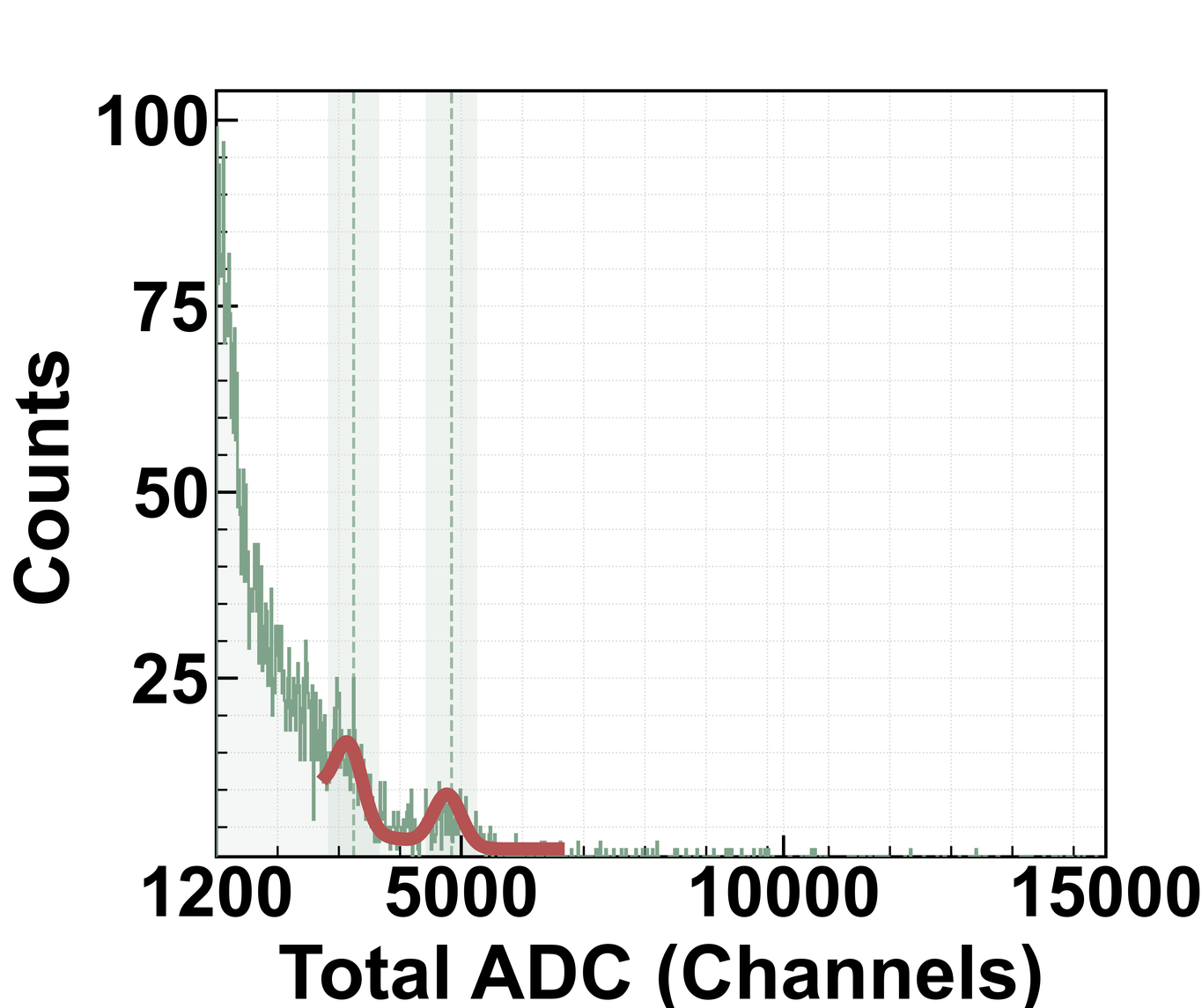}
        \caption{}
        \label{fig:radio_spec_lu_det2}
    \end{subfigure}
    \caption{The measured energy spectrum from radioactive sources: (a-c) $^{137}$Cs, (d-f) $^{241}$Am, (g-i) $^{176}$Lu for CH0, CH1 and CH2 in order, see text for details.}
    \label{fig:radio_spec}
\end{figure}

We use two radioactive sources, $^{241}$Am (59.5 keV) and $^{137}$Cs (662 keV), to calibrate the detector's response at low energy and around MeV energy range. The radioactive sources are placed closed to each detector and the detector is running in the calibration mode during the data taking. The energy spectra from one typical pixel for three detectors are shown in Figure~\ref{fig:radio_spec_cs_det0} through Figure~\ref{fig:radio_spec_am_det2}. After obtaining the ADC spectrum, a standard Gaussian function was used to fit it and extract the full absorption peaks.

In addition to the above two external radioactive sources, the LYSO scintillator which builds the CH0 contains the natural radioisotope $^{176}$Lu. This intrinsic radioactive source $^{176}$Lu could produce the characteristic gamma-rays (88 keV, 202 keV and 307 keV) associated with its $\beta$-decay \cite{AlvaSanchez2018}. Therefore, we make full use of these gammas to calibrate the detectors at more energy points. The energy spectra for three detectors in ADC are shown in Figure~\ref{fig:radio_spec_lu_det0} through Figure~\ref{fig:radio_spec_lu_det2}. The peaks are remarkably pronounced in CH0 and clearly distinguishable in CH1 and CH2.

A simulation with Geant4 was conducted to study the expected energy spectrum from the $\beta$-decay of $^{176}$Lu in LYSO. The $^{176}$Lu isotopes are uniformly distributed in CH0 and the energy depositions in CH0, CH1 and CH2 are derived. The $^{176}$Lu decay process consists of a beta emission accompanied by three prompt cascade gamma-rays with the energy of 88 keV, 202 keV and 307 keV. Inside CH0, the coincidence of the cascade gamma-rays produces the pronounced the peaks at 290 keV (sum of 88 keV and 202 keV), 395 keV (sum of 88 keV and 307 keV) and 597 keV (sum of 88 keV, 202 keV and 307 keV). These results are consistent with Ref.~\cite{AlvaSanchez2018}. In addition, the gamma-rays of 202 keV and 307 keV could escape from CH0 but then are observed by CH1 and CH2.

To fit the measured energy spectrum and extract the ADC values of the gamma peaks, we employ a function that convolves a continuous probability density function (PDF) of beta decay from the Geant4 simulations and an energy-dependent Gaussian function which represents the detector's energy resolution for each true gamma~\cite{AlvaSanchez2018}. This approach fully accounts for the True Coincidence Summing effect arising from the simultaneous energy deposition of the continuous beta and the cascade gamma-rays. The detailed procedure is described as follows.

Assuming $P(E)$ is the true energy spectrum from the simulation which completely preserves the continuous $\beta$ spectrum, the cascade $\gamma$-rays (88~keV, 202~keV and 307~keV), and their physical coincidence, but the energy is not yet smeared by the detectors. Let $R(adc, E)$ be the detector response function, the measured energy spectrum $S(adc)$ can be modeled below:
\begin{equation}
\label{eq:lu-fit}
S(adc) = A \int_{0}^{\infty} P(E) R(adc,~E) dE + C
\end{equation}
where $A$ is the normalization factor and $C$ is the constant background. The response function $R(adc, E)$ is assumed to be energy-dependent Gaussian:
\begin{equation}
R(adc, E) = \frac{1}{\sqrt{2\pi}\sigma(E)} \exp \left[ -\frac{\left(adc - g(E)\right)^2}{2\sigma^2(E)} \right]
\end{equation}
where $g(E)$ is a linear function that maps the physical energy deposition and the readout ADC value (details are discussed in Section~\ref{sec:calibration}), serving to directly align the energy domain of the simulated PDF with the ADC domain of the experimental data; $\sigma(E)$ is energy dependent and characterizes the spectrum spread at each reference gamma energy. 

The fitted results for one typical pixel of CH0 (virtual pixels which are manually divided for non-uniformity calibration, see more discussions in Section~\ref{sec:nonuniformity}) are shown in Figure~\ref{fig:radio_spec_lu_det0} where we derived the ADC values of the coincident gamma-ray peaks (290 keV, 395 keV and 597 keV). The energy peaks in CH1 and CH2 differ CH0 according to the MC simulation. By fitting the spectrum in Figure~\ref{fig:radio_spec_lu_det1} and \ref{fig:radio_spec_lu_det2}, the gamma-rays peaks (202 keV and 307 keV) were extracted for CH1 and CH2. These characteristic gammas are implemented in building the detector's energy response function which will be discussed in following. Moreover, the intrinsic gammas from $^{176}$Lu in CH0 will serve as the sources for the real-time self-calibration when the Compton camera is operated in orbit.

\subsection{Spatial Non-uniformity Calibration}
\label{sec:nonuniformity}
As described in Section~\ref{sec:design}, we use the capacitive multiplexing circuit to read out the signals from SiPMs which encodes an amount of SiPM outputs to four positional signals (PQRS). In this scenario, the position and energy for each event are convolved. Furthermore, consider the scintillator homogeneity, fabrication precision, photo-sensor's quantum efficiency, as well as the coupling between the scintillator and the SiPMs, the detector's response to the deposited energy is expected to be position dependent. Therefore we conduct the positional calibration to address this spatial non-uniformity effect.

Specifically, the positional calibration is performed by generating Flood-Maps and ID-Maps \cite{YaoZhiyang2023,Yang2026}. For CH1 and CH2, a radioactive source is placed at various positions in order to ensure all the crystal pixels can be fired by gammas. By analyzing the peak-to-valley ratio of the reconstructed positions according to Eq.~\ref{eq:position12}, the Flood-Map is obtained and then further regularized into a standard ID-Map. The procedure for CH0 is similar but we manually divided the detector into virtual pixels since it uses a monolithic crystal bulk. Practically, it is divided into 7$\times$7 grids. A radioactive source is placed at the center of each grid for data taking, and then 49 calibration data sets are collected. Each individual region is distinguished based on the peak-to-valley ratio between the reconstructed source positions using Eq.~\ref{eq:position0}. {In addition, to align all three detector layers, a 2D interpolation is employed to further refine the 7$\times$7 grids to 14$\times$14 grids which is used to build the final ID-Map for CH0. The center coordinates of the grids from the ID-Maps are used as the final reconstructed positions for each hit.
\begin{figure}[htbp]   
    \centering
    \begin{subfigure}[t]{0.3\columnwidth}
        \includegraphics[width=\linewidth]{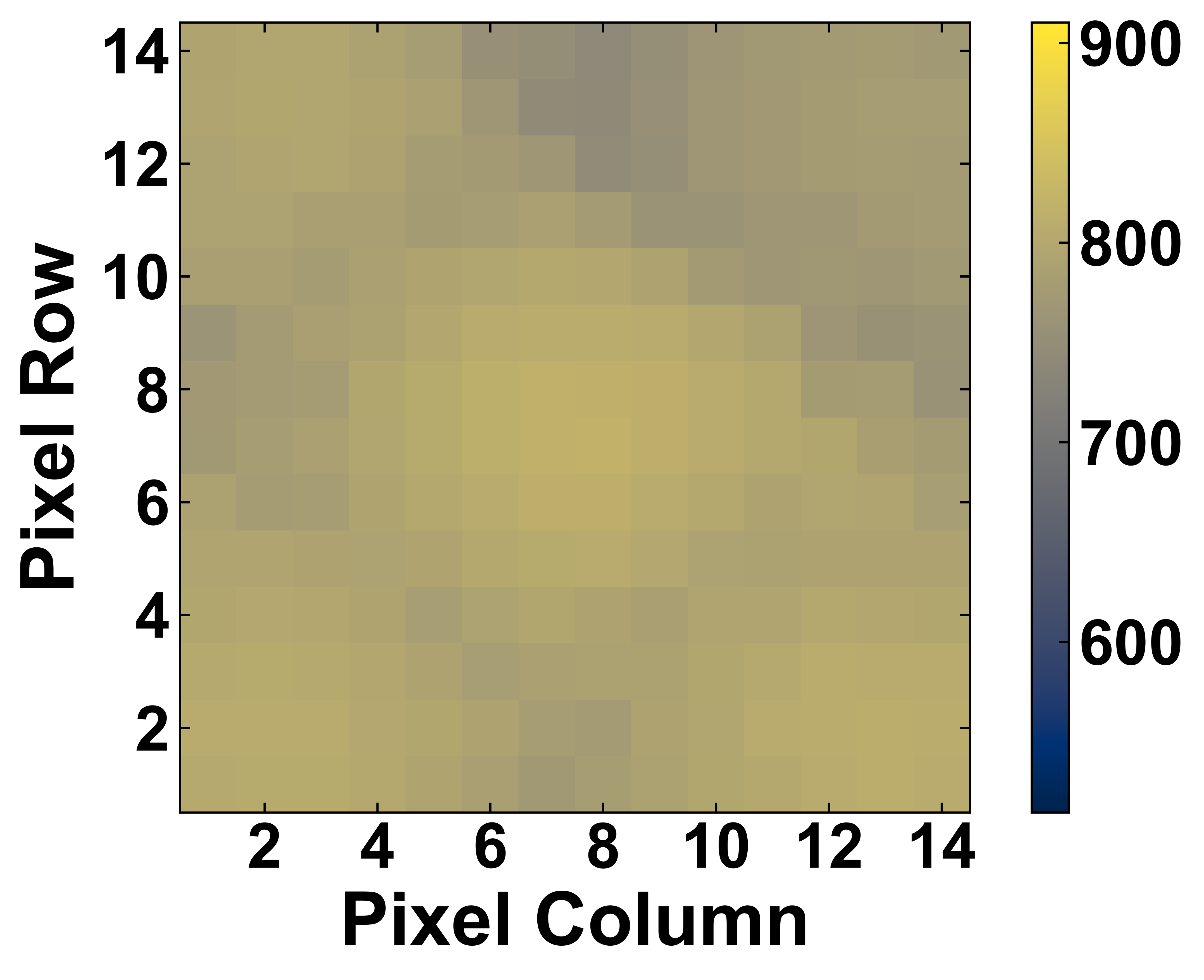}
        \caption{}
        \label{fig:nonuniform-2d-det0}
    \end{subfigure}
    \hfill
    \begin{subfigure}[t]{0.3\columnwidth}
        \includegraphics[width=\linewidth]{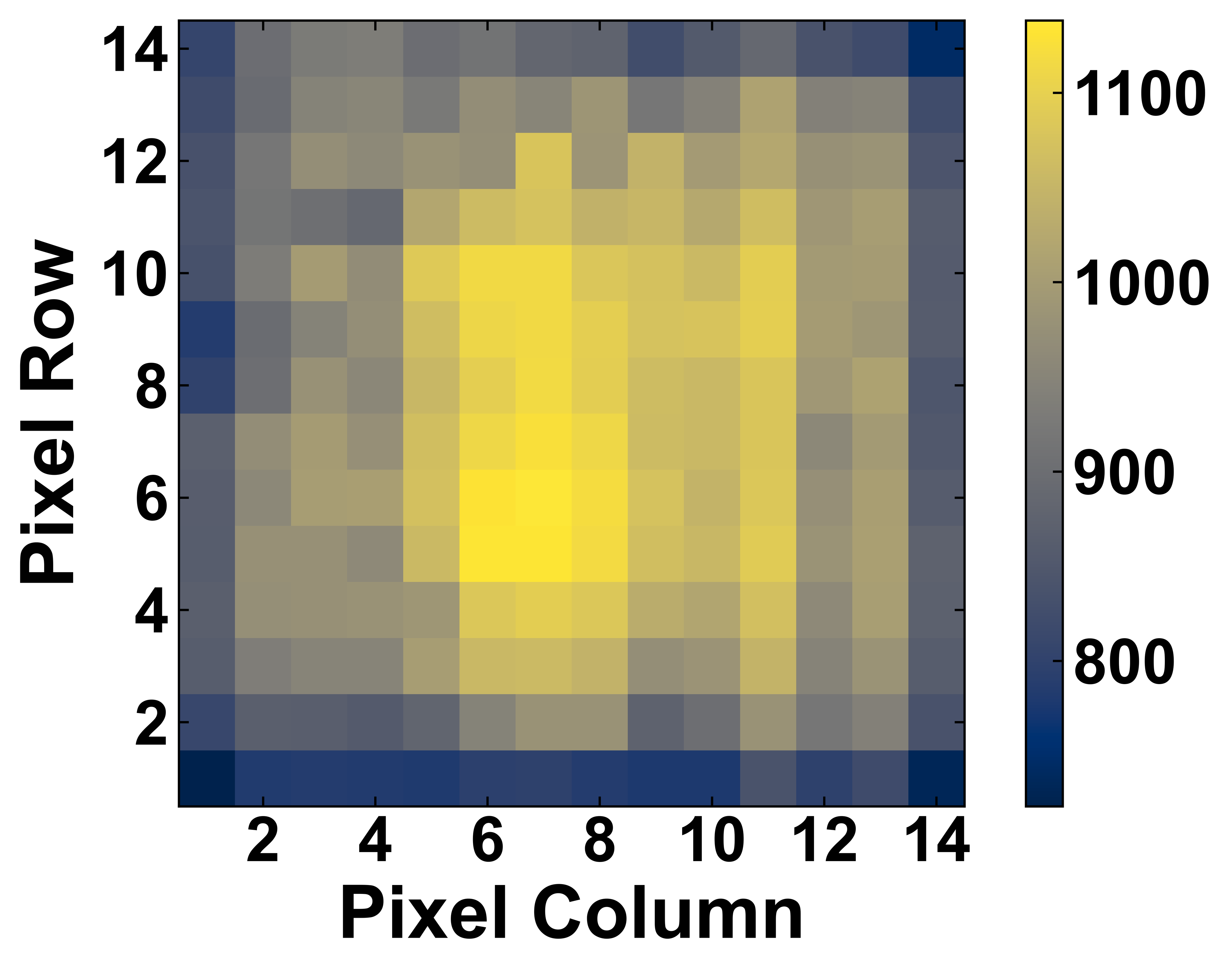}
        \caption{}
        \label{fig:nonuniform-2d-det1}
    \end{subfigure}
    \hfill
    \begin{subfigure}[t]{0.3\columnwidth}
        \includegraphics[width=\linewidth]{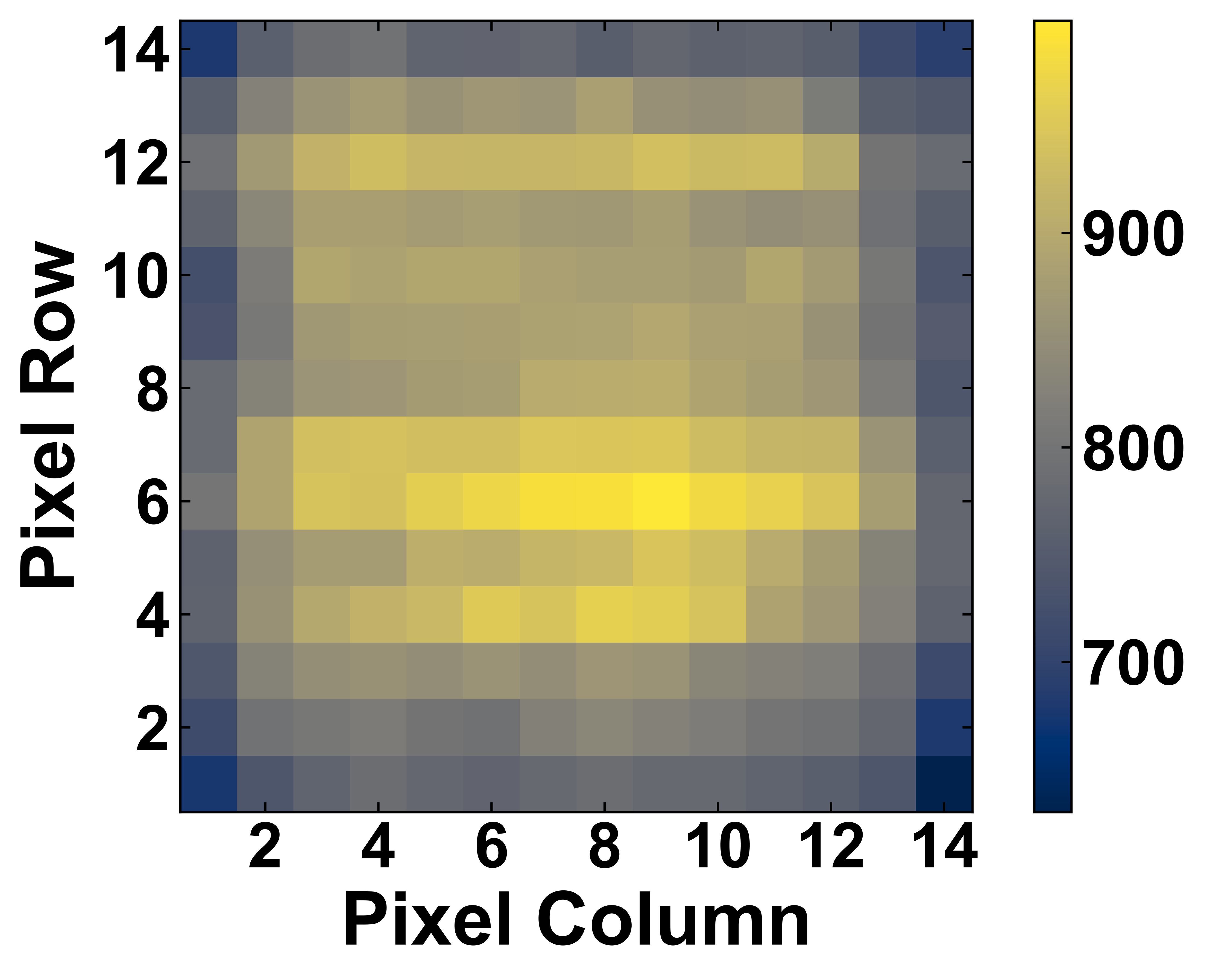}
        \caption{}
        \label{fig:nonuniform-2d-det2}
    \end{subfigure}
    \par
    \begin{subfigure}[t]{0.3\columnwidth}
        \includegraphics[width=\linewidth]{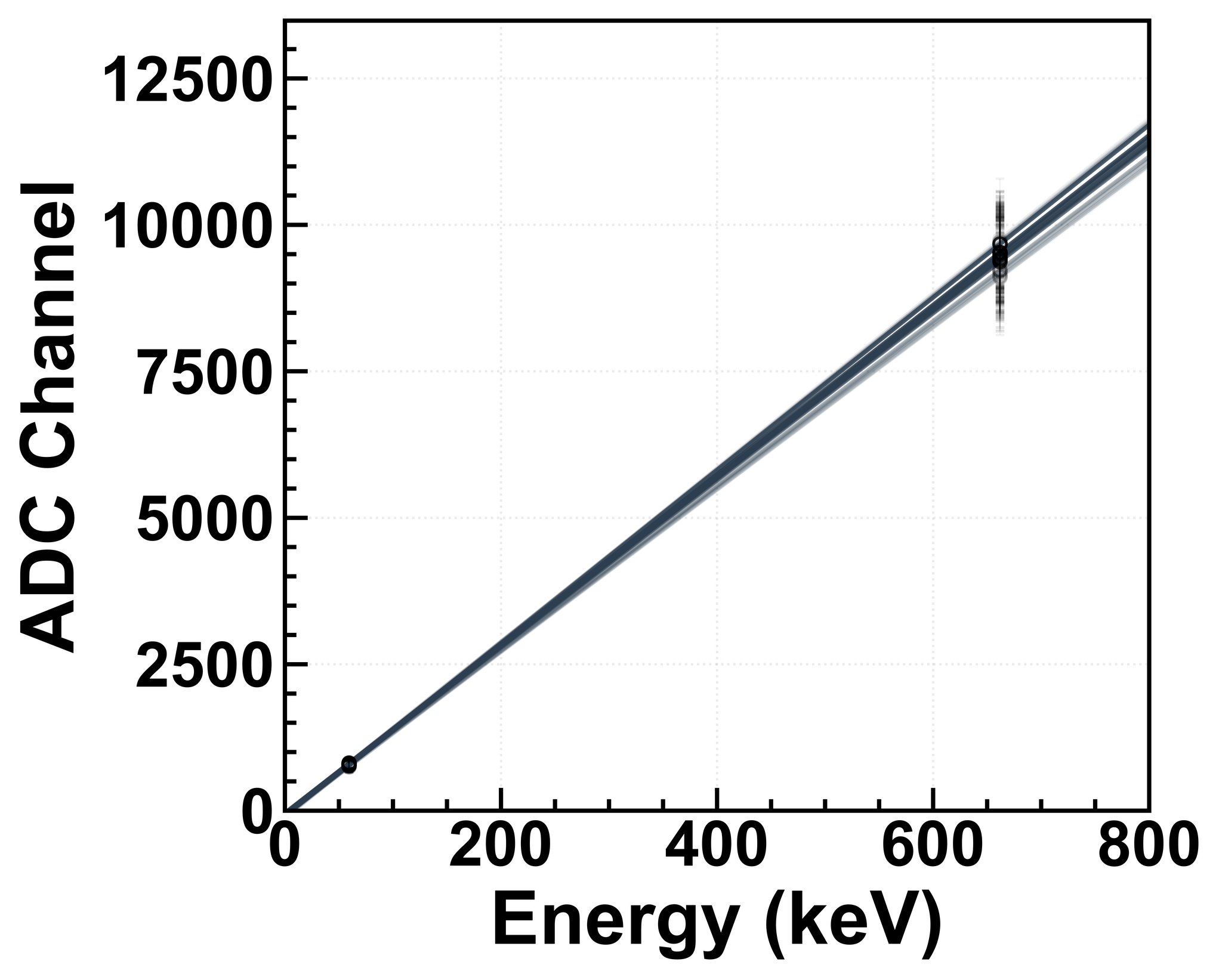}
        \caption{}
        \label{fig:nonuniform-det0}
    \end{subfigure}
    \hfill
    \begin{subfigure}[t]{0.3\columnwidth}
        \includegraphics[width=\linewidth]{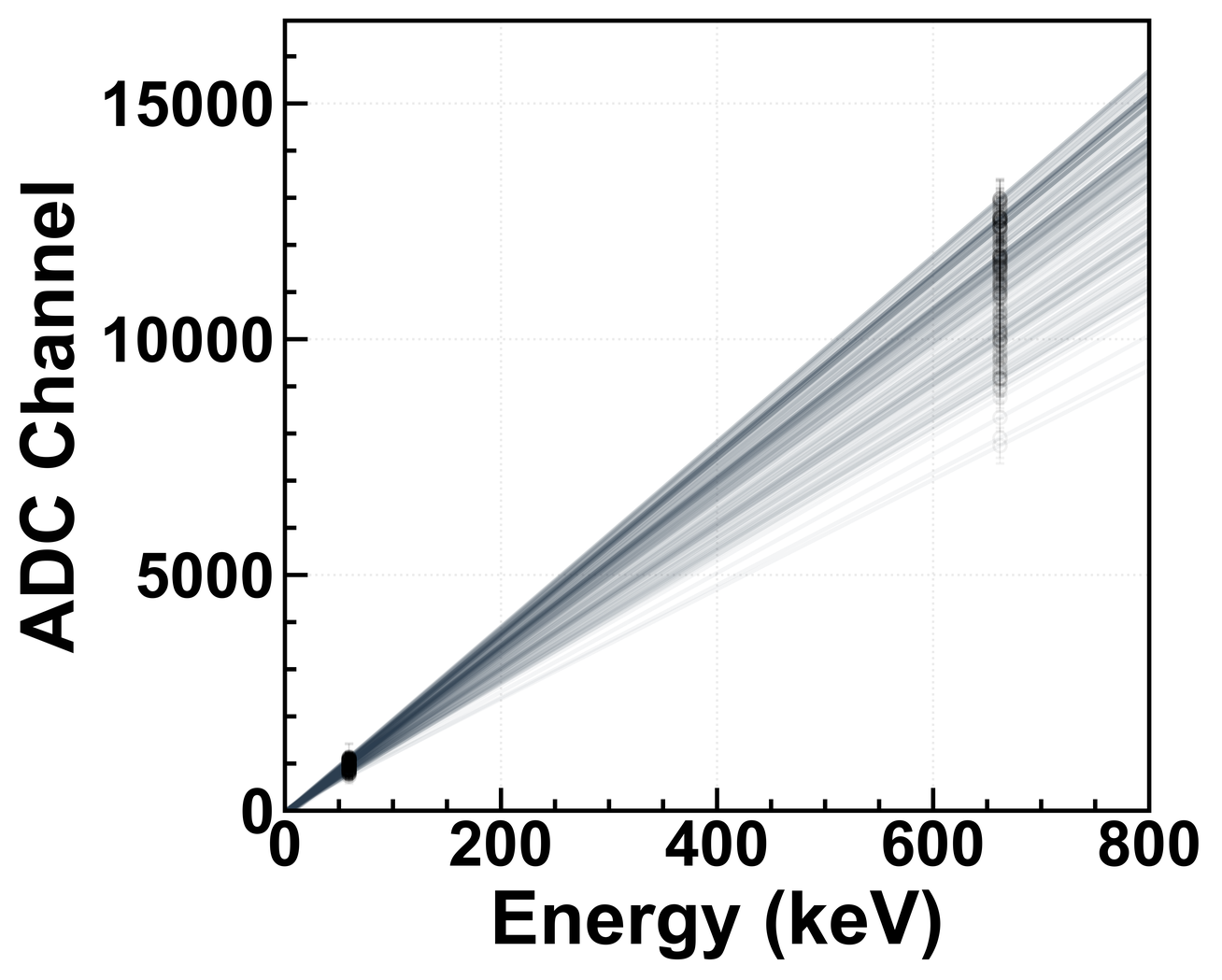}
        \caption{}
        \label{fig:nonuniform-det1}
    \end{subfigure}
    \hfill
    \begin{subfigure}[t]{0.3\columnwidth}
        \includegraphics[width=\linewidth]{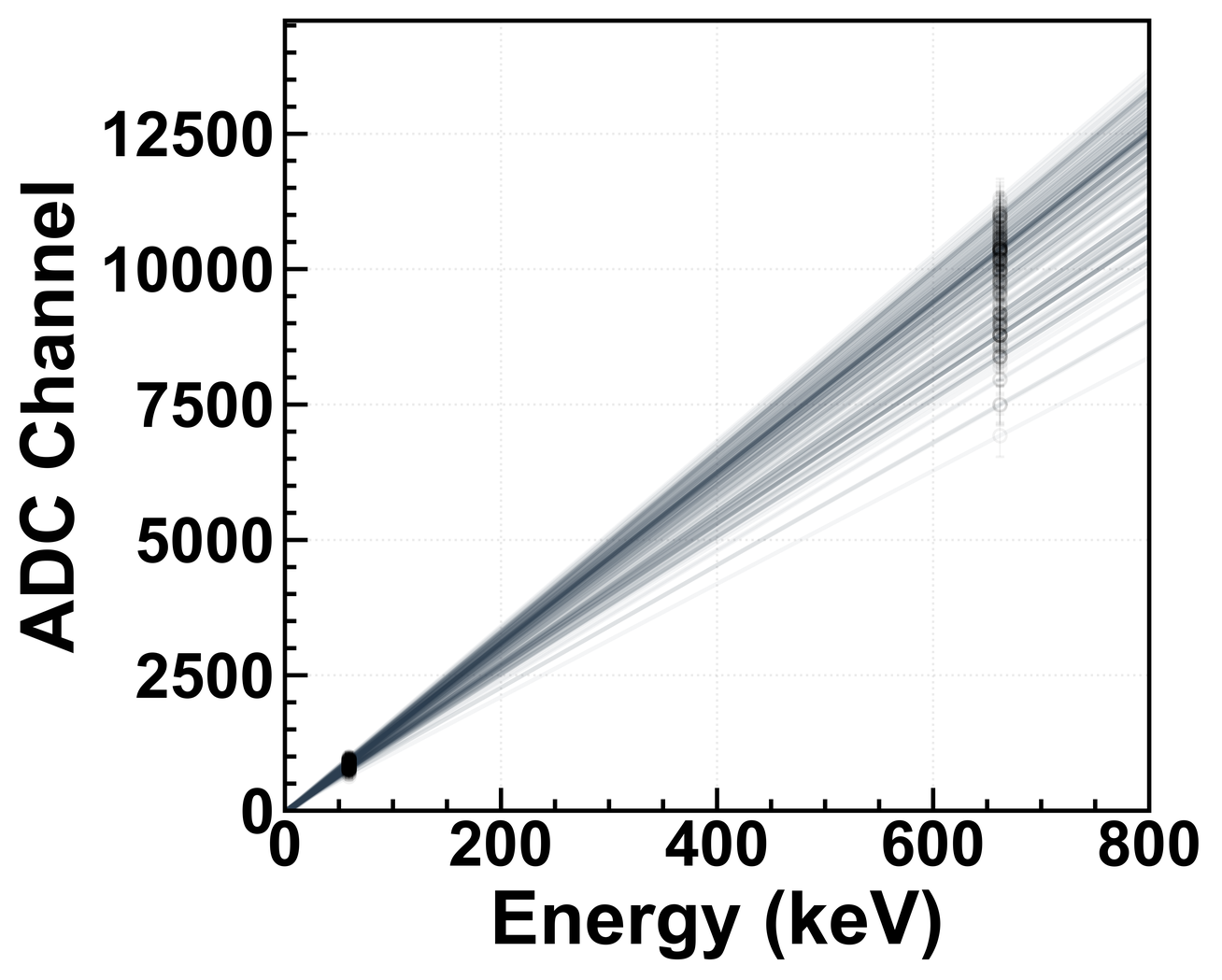}
        \caption{}
        \label{fig:nonuniform-det2}
    \end{subfigure}
    \caption{Non-uniformity calibration results: the upper heat maps are the ADC values of $^{241}$Am peak for CH0 (a), CH1 (b) and CH2 (c) where the color represents the ADC values. The lower plots are the energy scale function for individual detector pixel for CH0 (d), CH1 (e) and CH2 (f).}
    \label{fig:nonuniform}
\end{figure}

With the calibration data of $^{241}$Am, the mean ADC value of the full energy absorption peak (59.5 keV) are derived for each individual detector pixel. Figure~\ref{fig:nonuniform-2d-det0}, ~\ref{fig:nonuniform-2d-det1} and ~\ref{fig:nonuniform-2d-det2} shows the corresponding heat map for CH0, CH1 and CH2, respectively. We repeated such calibration with $^{137}$Cs radioactive source and produced the similar mapping. In addition, assuming a linear detector response and zero ADC value for zero energy deposition, we fitted the signal in ADC as the function of the energies (0 keV, 59.5 keV and 662 keV) for each individual pixel and the results are shown in Figure~\ref{fig:nonuniform-det0}, ~\ref{fig:nonuniform-det1} and ~\ref{fig:nonuniform-det2}. It's concluded that the detector's response to energy is very linear in each individual pixel while the response function is obviously varied at different positions across the detector layer. The non-uniformity at 662 keV is around 10\% for CH1 and CH2, and $<$5\% for CH0. The position-dependent energy response mapping for each detector layer is built and will be used for event-by-event correction in the physics data analysis.

\subsection{Energy Linearity \& Resolution}
After applying the spatial non-uniformity correction, the energy spectrum in ADC for each detector layer was obtained. Using the fitting algorithms discussed previously, the mean ADC value as well as its spread at each energy point were extracted, and the results are shown in Figure~\ref{fig:energy_scale}.

Assuming a linear response within the energy range of interest, we define the below function to describe the relationship between the measured signal, denoted as $\textit{S(ADC)}$ and the deposited energy for each detector: 
\begin{equation}
S_i(ADC) = a_i \cdot E~(keV) + b_i
\label{eqa:linear}
\end{equation}
where $i$ labels the detector layer index, $S$ is the sum of PQRS readouts from the electronics and $E$ is the deposited energy in keV; $a$ is the energy conversion factor and $b$ is a constant in ADC. The fitting results for CH0, CH1 and CH2 are shown as the lines in Figure~\ref{fig:energy_scale}. Table~\ref{tab:energy_scale} summarizes the fitted values of $a$ and $b$ coefficients for the three detector layers. The energy non-linearity is defined as the fractional difference between the actual readout ADC and the fitted value. Overall the detectors performed a good linear response within the MeV energy range. At the energy of 662 keV, the non-linearity is 1.1\%, 0.8\% and 0.5\% for CH0, CH1 and CH2, respectively. This good linearity in energy response would be essential for the gamma-ray measurement using the multi-Compton-scattering detection.
\begin{table}[htbp]
    \centering
    \caption{Summarized parameters of energy scale for each detector.}
    \label{tab:energy_scale}
    \begin{tabular}{c|c|c}
     \hline 
         & \multirow{2}*{a [ADC/keV]}  & \multirow{2}*{b [ADC]} \\
         &                             &    \\
    \hline
     Detector 0/CH0  & 14.78 $\pm$ 0.23    & 36.43 $\pm$ 95.68    \\
     Detector 1/CH1  & 19.16 $\pm$ 0.39    & -49.36 $\pm$ 133.58  \\
     Detector 2/CH2  & 15.83 $\pm$ 0.21    & -20.58 $\pm$ 69.81   \\
    \hline
    \end{tabular}
\end{table}

The fractional energy resolution at each energy point was calculated with the mean ADC value and its spread from the above spectrum fitting. The resolutions as the function of energies for each detector layer are shown as the points in Figure~\ref{fig:energy_res}. The error bar of each point is the uncertainty from the spectrum fitting. After the spatial non-uniformity correction, the energy resolution at 662 keV reaches to 3.3\%, 3.7\%, 3.9\% for CH0, CH1 and CH2. In addition, a simple fitting with the function (\textit{f}($\frac{\sigma}{E}) = \frac{\alpha}{\sqrt{E}} + \beta$) is applied, and the results are shown as the lines on top of the data points in Figure~\ref{fig:energy_res}. The energy-dependent resolutions basically follow the statistical fluctuations as expected.
\begin{figure}[htbp]   
	\centering
   \begin{subfigure}[t]{0.45\columnwidth}
        \includegraphics[width=\linewidth]{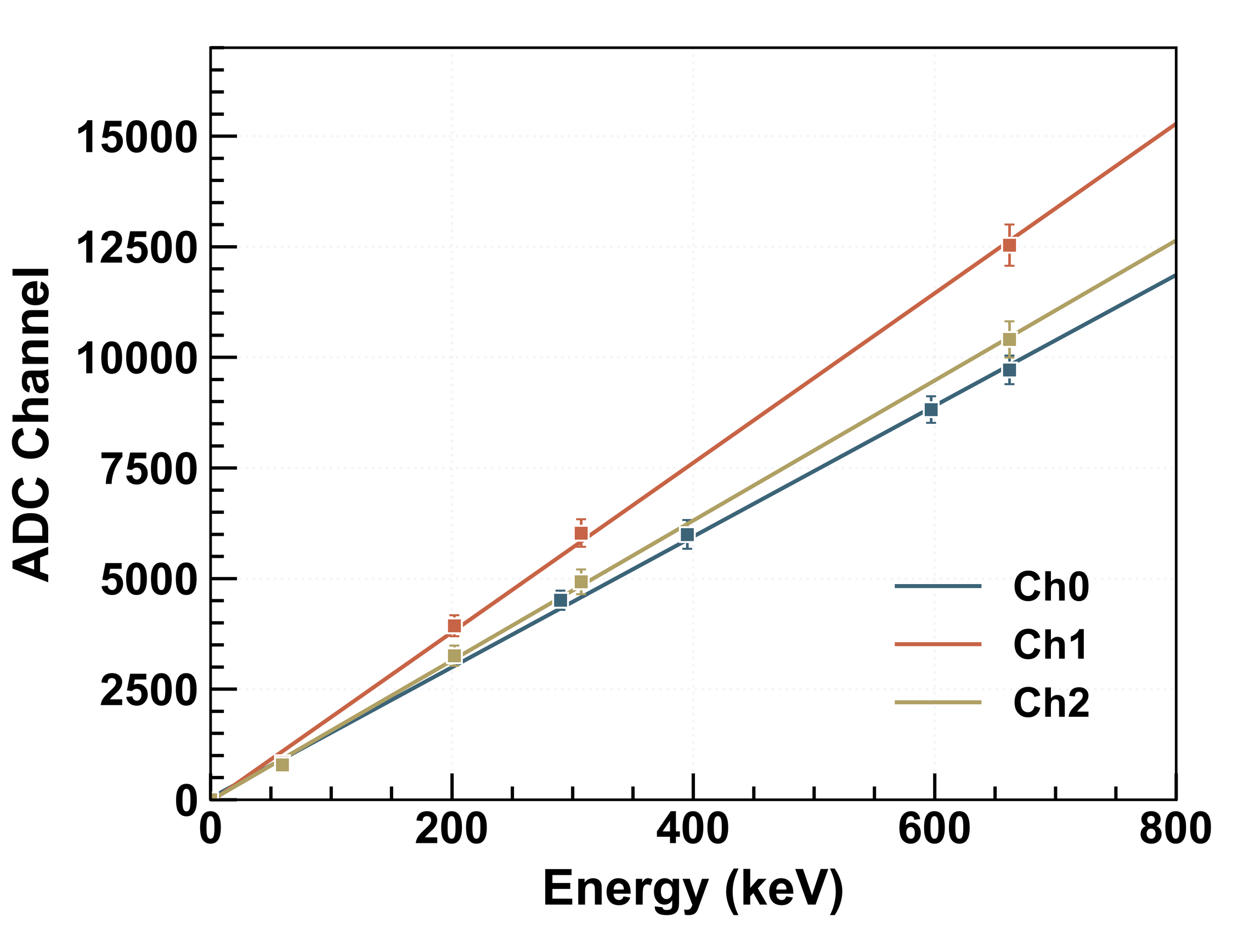}
        \caption{}
        \label{fig:energy_scale}
    \end{subfigure}
    \hspace{0.25cm}
    \begin{subfigure}[t]{0.45\columnwidth}
        \includegraphics[width=\linewidth]{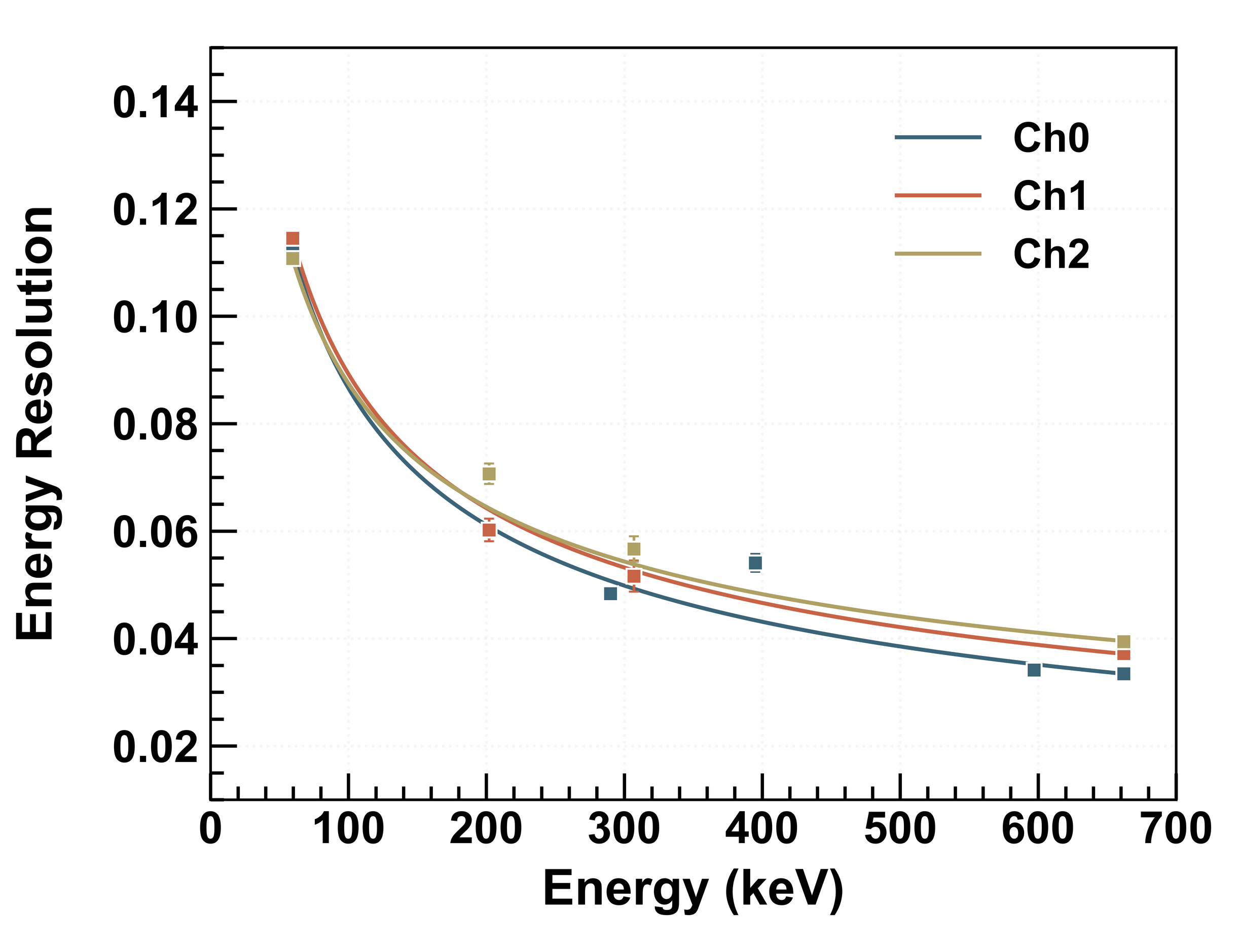}
        \caption{}
	   \label{fig:energy_res}
    \end{subfigure}
\caption{Calibrated energy linearity and resolutions. (a) Energy scale of each detector. (b) Fractional energy resolution as the function of energy for each detector, overlaid with fittings.} 
\label{}
\end{figure}

\section{Instrumentation Performances}
\label{sec:performance}
We present the detailed performances of the Compton camera at the system level in this section including the results of track reconstruction with cosmic muons and the systematic resolution in energy and angle for the MeV gamma-ray imaging. For the results discussed here, the data were taken when the Compton camera was set at the working mode (i.e. coincidence-trigger on the two scattering layers). 

\subsection{Muon Reconstruction}
The Compton camera was set to point up to the sky for taking the data of cosmic muons. The threshold of CH1 and CH2 were configured at 1000 DAC which was abound 427.6 keV and the threshold of CH0 was set at 500 DAC (1 DAC$\approx$0.25 mV), corresponding to 330 keV. The veto threshold was set as the upper limit of the electronics, in order to make full use of its dynamic range. The average trigger rate was $\sim$10 events per minute which was consistent with the expectation.

With $\sim$15h data taking, we collected $\sim$4000 valid muons defined as the events which passed through all three detector layers. Implementing the energy calibration and position correction as discussed in Section~\ref{sec:calibration}, the muon tracks and energy spectra were reconstructed. At the sea level, muons are very relativistic with the average kinetic energy of a few GeV, so in principle the cosmic muons travel through the three detector layers straightly~\cite{PDG2024}. Figure~\ref{fig:mu_track} shows the reconstructed muon tracks from the Compton camera. In addition, we fitted the muon hits on three layers with a straight line using the least squares method and calculated the deviations between the actual reconstructed positions and the fitted tracks. The results are shown in Figure~\ref{fig:mu_track_err}. The Compton camera performed the excellent position reconstruction with an accuracy of $\sim$1 mm for all the three layers which is critical for imaging gamma-rays using the multi-Compton-scattering technique. These results also well validate the non-uniformity correction and the track reconstruction for charged particles.
\begin{figure}[htbp]
    \centering
    \begin{subfigure}[t]{0.4\columnwidth}
        \includegraphics[width=\linewidth]{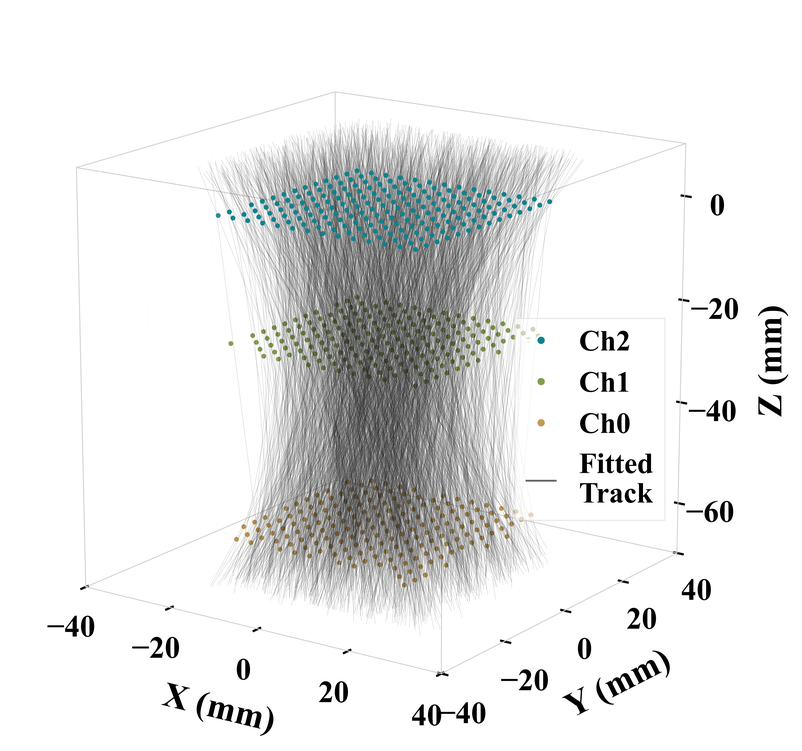}
        \caption{}
        \label{fig:mu_track}
    \end{subfigure}
    \hspace{1cm}
    \begin{subfigure}[t]{0.45\columnwidth}
        \includegraphics[width=\linewidth]{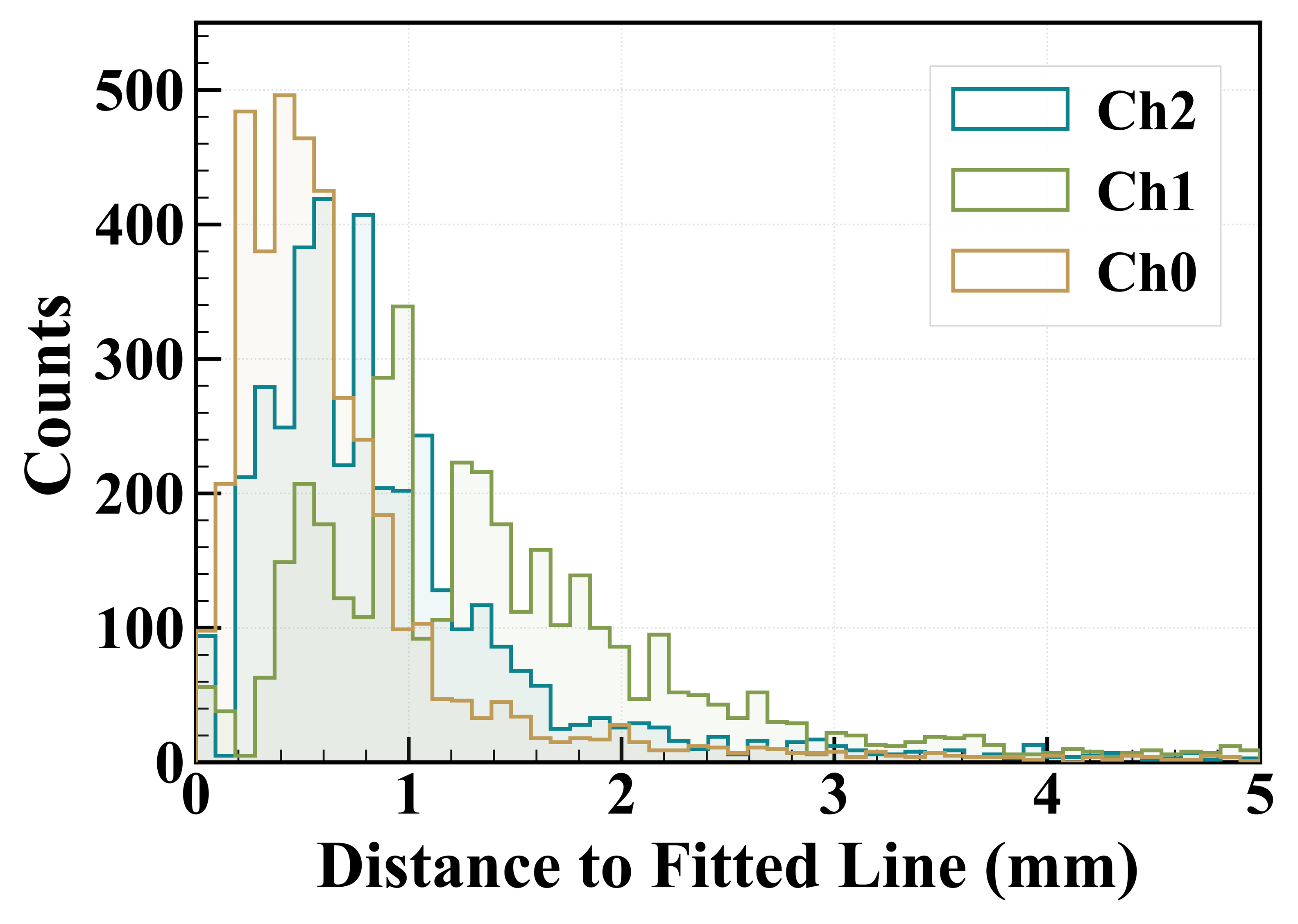}
        \caption{}
        \label{fig:mu_track_err}
    \end{subfigure}
    \begin{subfigure}[t]{0.3\columnwidth}
        \includegraphics[width=\linewidth]{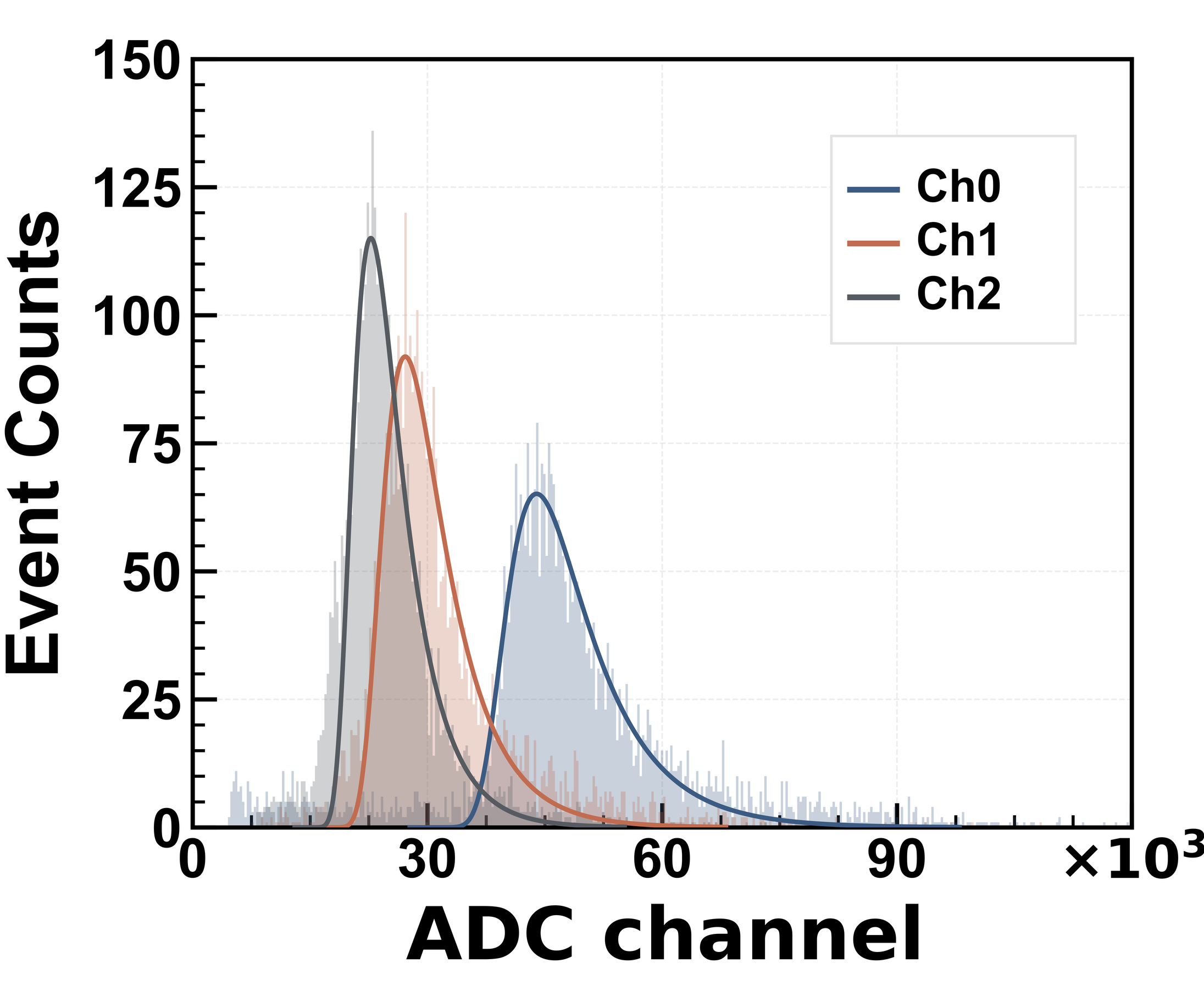}
        \caption{}
        \label{fig:mip_spec_raw}
    \end{subfigure}
    \hspace{0.5cm}
    \begin{subfigure}[t]{0.3\columnwidth}
        \includegraphics[width=\linewidth]{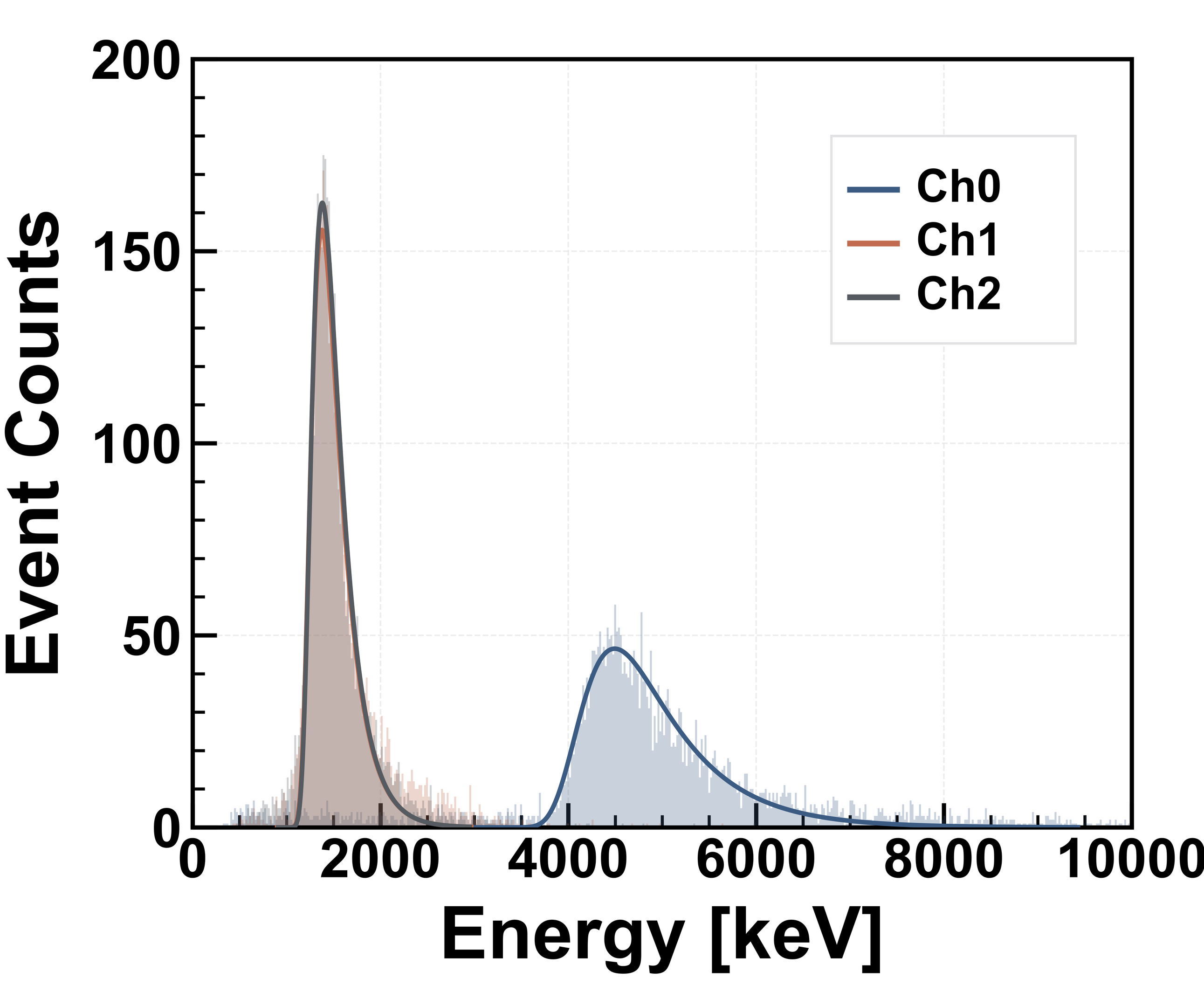}
        \caption{}
        \label{fig:mip_spec_position}
    \end{subfigure}
    \hspace{0.5cm}
    \begin{subfigure}[t]{0.3\columnwidth}
        \includegraphics[width=\linewidth]{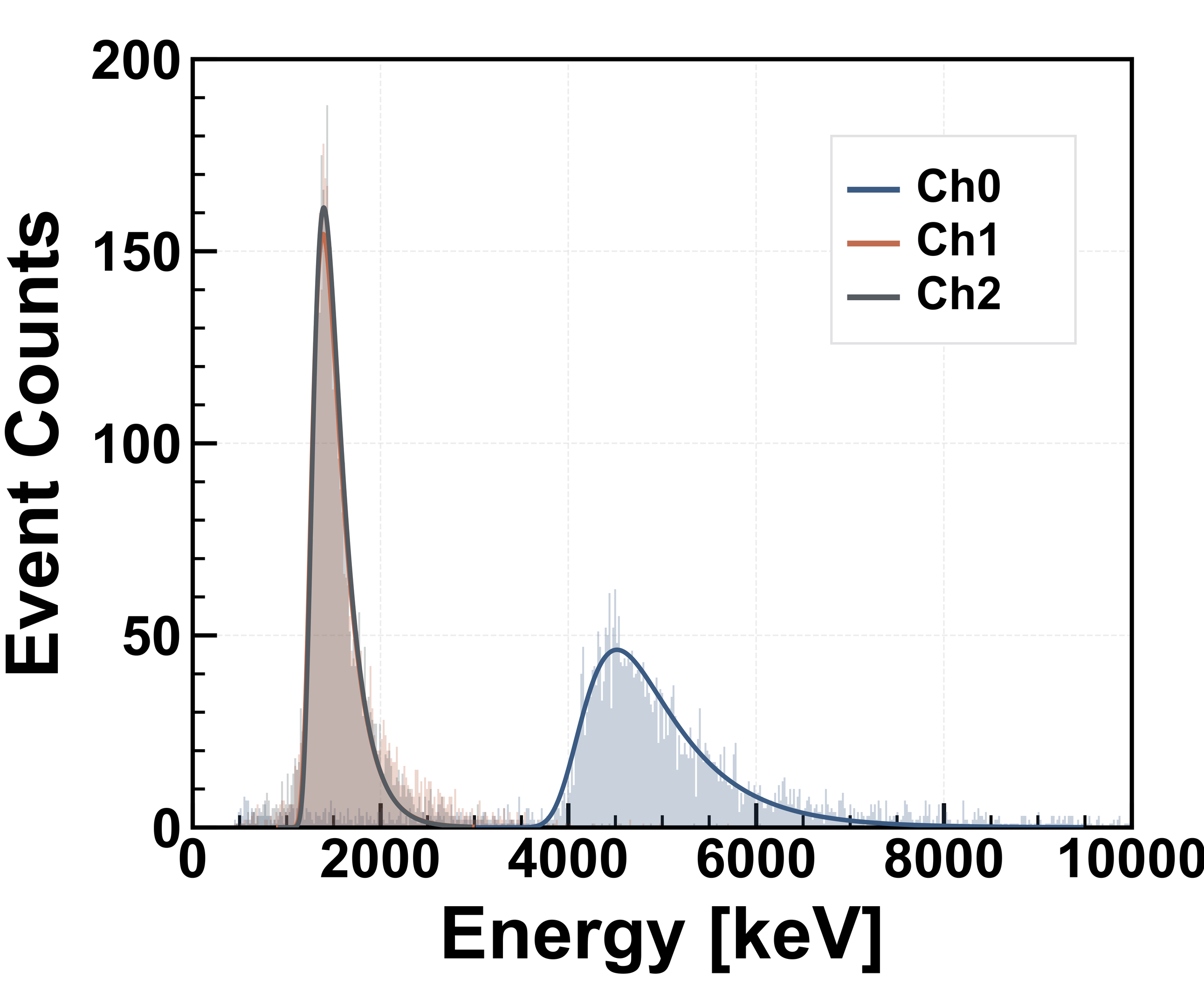}
        \caption{}
        \label{fig:mip_spec_final}
    \end{subfigure}
    \caption{Measurement of cosmic muons: (a) hits on each detector layer associated with the reconstructed tracks; (b) the distance between the actual hit position and the reconstructed tracks; (c) MIP spectrum for each detector layer in raw data; (d) MIP spectra after the position correction; (e) MIP spectra after the correction of position and incident angle.}
    \label{fig:muons}
\end{figure}

Furthermore, we analyzed the data to reconstruct the MIP (Minimum Ionization Particle) energy spectrum. Figure~\ref{fig:mip_spec_raw} shows the raw MIP spectrum from each detector while Figure~\ref{fig:mip_spec_position} shows the spectra after the non-uniformity correction. In addition, muons hit the detectors at various incident angles \cite{PDG2024,Benaglia2016} which would cause different lengths of passage paths inside the detector and smears the energy deposition consequently. Relying on the track reconstruction discussed previously, we calculated the incident angle of muons and applied the angular correction event by event. The results are shown in Figure~\ref{fig:mip_spec_final}. We fitted the spectrum with a standard Landau function and extracted the MPV (Most Probable Value) and resolution of MIP. Table~\ref{tab:mip} summarizes the fitted results. The MPV of MIPs is 4.86$\pm$0.26 MeV, 1.63$\pm$0.09 MeV and 1.56$\pm$0.08 MeV for CH0, CH1 and Ch2, respectively, which are well consistent with the simulations. The energy resolutions of three detectors are obviously improved after the spatial non-uniformly and incident angle correction, and reach to $\sim$5.5\% for MIPs.
\begin{table}[htbp]
    \centering
    \caption{Energy resolution of MIP for three detector layers. The MPV and $\sigma$ is in unit of ADC for the raw spectrum and keV for the spectra after the position or incident angle correction.} 
    \label{tab:mip}
    \begin{tabular}{c|c|c|c|c}
    \hline
    \multirow{2}{*}{Chan.} & \multirow{2}{*}{Condition} & MPV &$\sigma$ & \multirow{2}{*}{Resolution} \\
                          &  & [ADC or keV] &[ADC or keV] & \\
    \hline
    \multirow{3}{*}{Ch0} & Raw &47870.7 &2924.5 &6.1\% \\
                         & Posit. Correct. &4802.6 &263.8 &5.5\% \\
                         & Track Correct. &4862.4 &262.4 &5.4\% \\
    \hline
    \multirow{3}{*}{Ch1} & Raw &31835.8 &2129.2 &6.7\% \\
                         & Posit. Correct. &1599.6 &90.6 &5.7\% \\
                         & Track Correct. &1630.9 &91.1 &5.6\% \\
    \hline
    \multirow{3}{*}{Ch2} & Raw &25150.2 &1574.8 &6.3\% \\
                         & Posit. Correct. &1534.1 &81.7 &5.3\% \\
                         & Track Correct. &1563.4 &85.3 &5.4\% \\
    \hline
    \end{tabular}
\end{table}

\subsection{Gamma-Ray Source Imaging}
A $^{137}$Cs radioactive source was used to evaluate the instrumentation performance for the MeV gamma-ray imaging in the Compton-scattering mode. Experimental data were acquired in the working mode (coincidence trigger on the scattering layers CH1 and CH2). The threshold gate was set at [60, 1500] (DAC) which is roughly from 31 to 767 keV. The veto threshold is set well below the MIP energy in order to minimize the effect of cosmic muons. The coincidence trigger substantially suppressed the backgrounds from the radioactivity of $^{176}\mathrm{Lu}$ intrinsically contained in CH0 as well as the environmental noise. 

The angular resolution measure (ARM) is analyzed to characterize the angular response of the Compton camera~\cite{V.Schönfelder2000,Takeda2007ARM,Hosokoshi2019ARM}. By definition, the ARM indicates the reconstruction accuracy of the Compton scattering angle and can be calculated with below formula:
\begin{equation}
    \mathrm{ARM} = \theta_{\mathrm{E}} - \theta_{\mathrm{G}}.
    \label{eq:arm_definition}
\end{equation}
where $\theta_{\mathrm{E}}$ is the kinematic scattering angle which is determined from the measured energy depositions, and $\theta_{\mathrm{G}}$ the geometrical scattering angle that is derived from the reflected trajectories in the reconstruction.

For simplicity, in the analysis presented here, the $^{137}$Cs events with two coincident hits recorded in the first two detector layers (CH2 and CH1) were selected to calculate the ARM. The two interaction positions were denoted by $\mathbf{p}_{\mathrm{2}}$ and $\mathbf{p}_{\mathrm{1}}$, with the corresponding deposited energy of $E_{\mathrm{2}}$ and $E_{\mathrm{1}}$. Additionally, a cut on energy ($0.652\text{ MeV} \le E_{\gamma} \le 0.672\text{ MeV}$) was applied to the reconstructed energy to select the full absorption events of $^{137}$Cs. According to the Compton scattering kinematics, $\theta_{\mathrm{E}}$ was calculated with the below formula:
\begin{equation}
    \cos \theta_{\mathrm{E}} = 1 - m_e c^2 \left( \frac{1}{E_{\mathrm{1}}} - \frac{1}{E_{\mathrm{2}} + E_{\mathrm{1}}} \right),
    \label{eq:angular_reconstruction}
\end{equation}
where $m_e c^2\approx0.511$~MeV is the rest mass of the electron, and $E_{\gamma}=E_{1}+E_{2}$ represents the reconstructed energy of an incident gamma with being full absorbed. Independently, the geometrical scattering angle $\theta_{\mathrm{G}}$ was determined from the source position $\mathbf{s}$ which was prior measured and the coordinates of the two interactions:
\begin{equation}
    \cos \theta_{\mathrm{G}} = \frac{\overrightarrow{p_s p_2} \cdot \overrightarrow{p_2 p_1}}{|\overrightarrow{p_s p_2}| |\overrightarrow{p_2 p_1}|}.
    \label{eq:geometric_reconstruction}
\end{equation}

It is important to adopt a correct physical sequence of the successive interactions for the angular reconstruction using the technique of Compton imaging. A kinematic sequencing algorithm was developed to address this issue. For each two-hit event, the algorithm tests both two possible trajectory orders. If only one trajectory order yields $|\cos\theta_{\mathrm{E}}|\le1.0$, it is accepted. If both trajectory orders are valid, the algorithm resolves the directional ambiguity by selecting the sequence that minimizes the absolute ARM value ($|\theta_{\mathrm{E}} - \theta_{\mathrm{G}}|$). In addition, a back-scattering filter was applied to remove the events with too large scattering angles because such events likely suffered from large geometric uncertainties or low-energy noise from the environment.
\begin{figure}[htbp]   
\centering
\begin{subfigure}[t]{0.6\columnwidth} 
    \includegraphics[width=\linewidth]{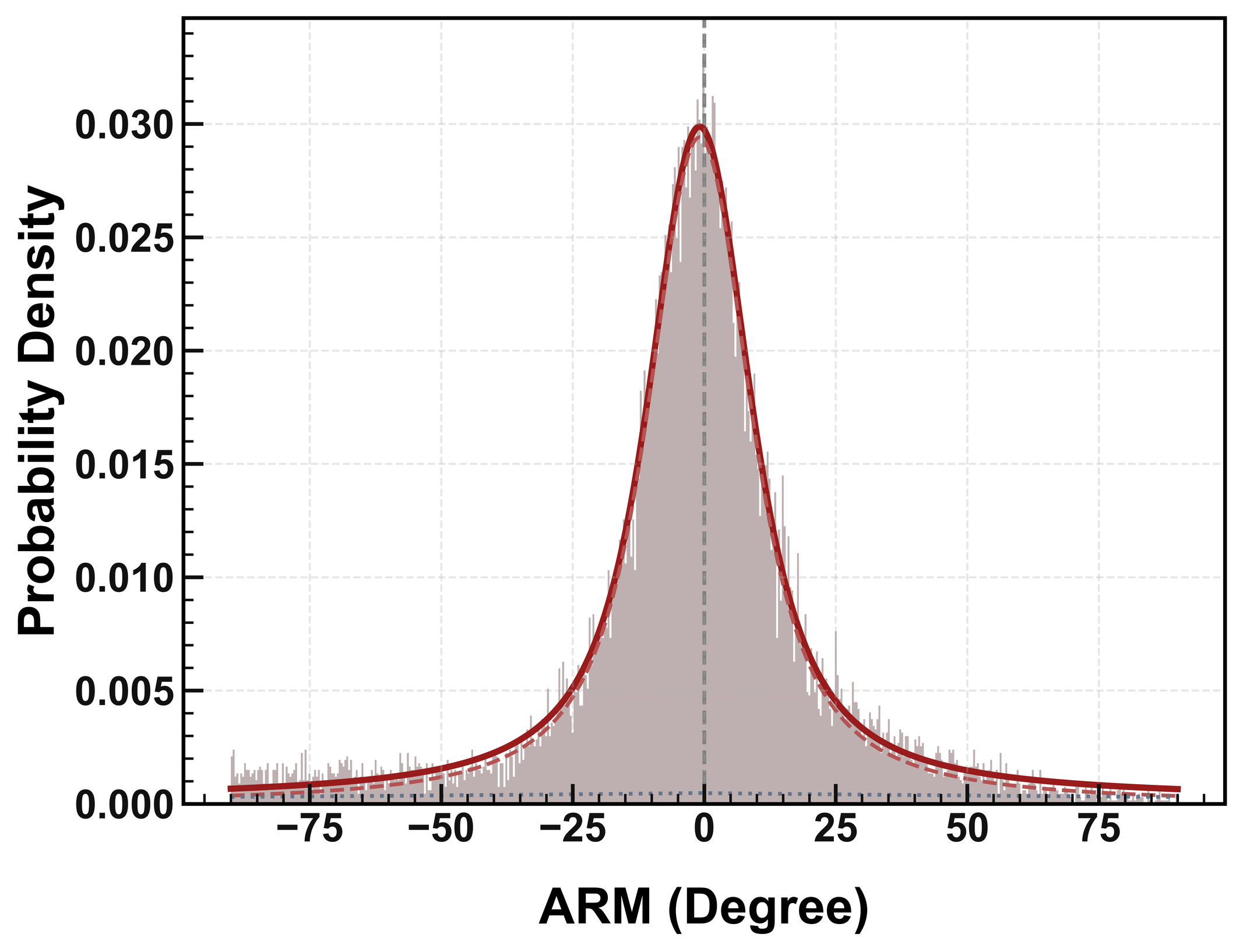}
    \caption{}
    \label{fig:arm}
\end{subfigure}
\par
\begin{subfigure}[t]{0.35\columnwidth}
    \includegraphics[width=\linewidth]{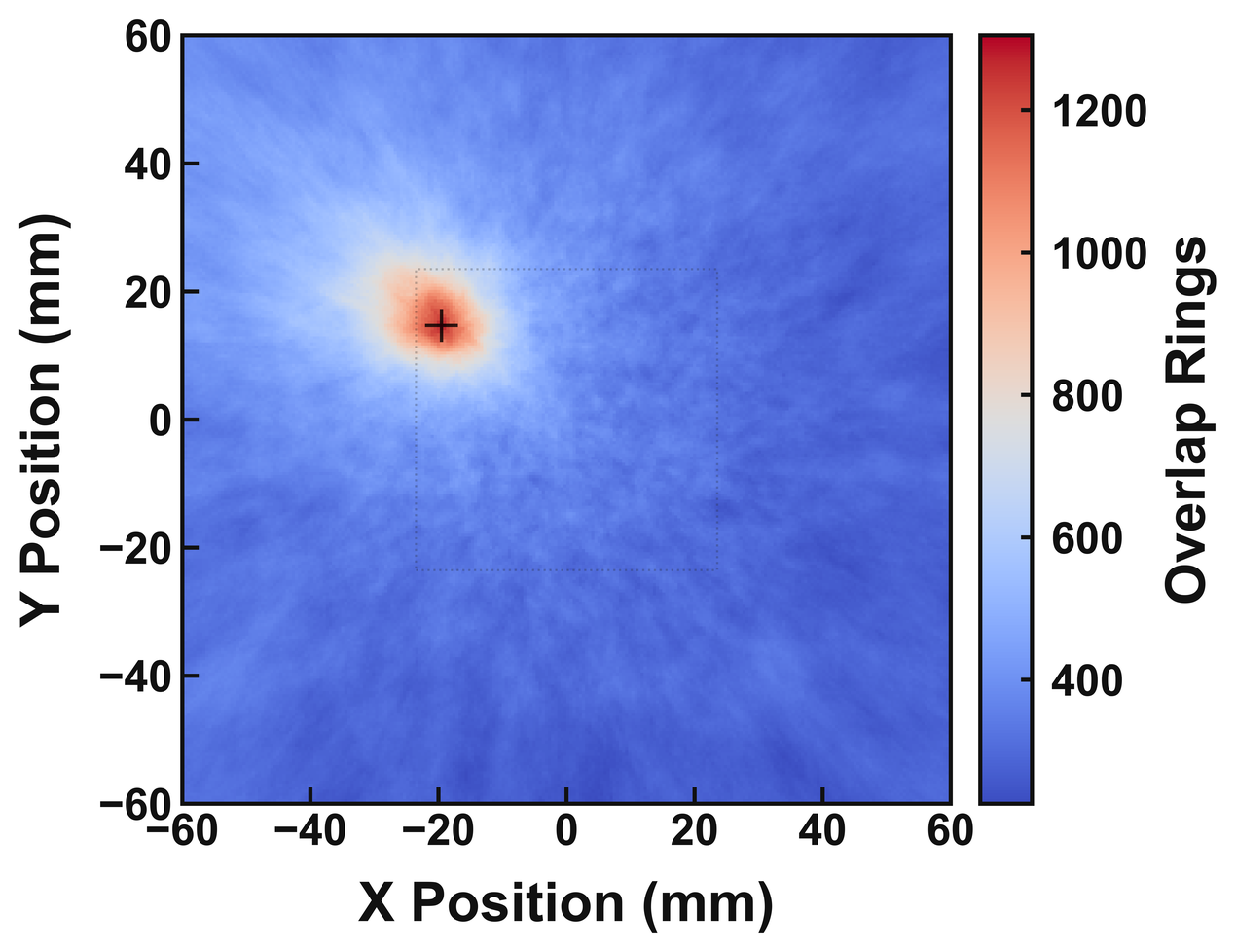}
    \caption{P$_{1}$ with SBP}
    \label{fig:SBPP1}
\end{subfigure}
\hspace{0.5cm}
\begin{subfigure}[t]{0.35\columnwidth}
    \includegraphics[width=\linewidth]{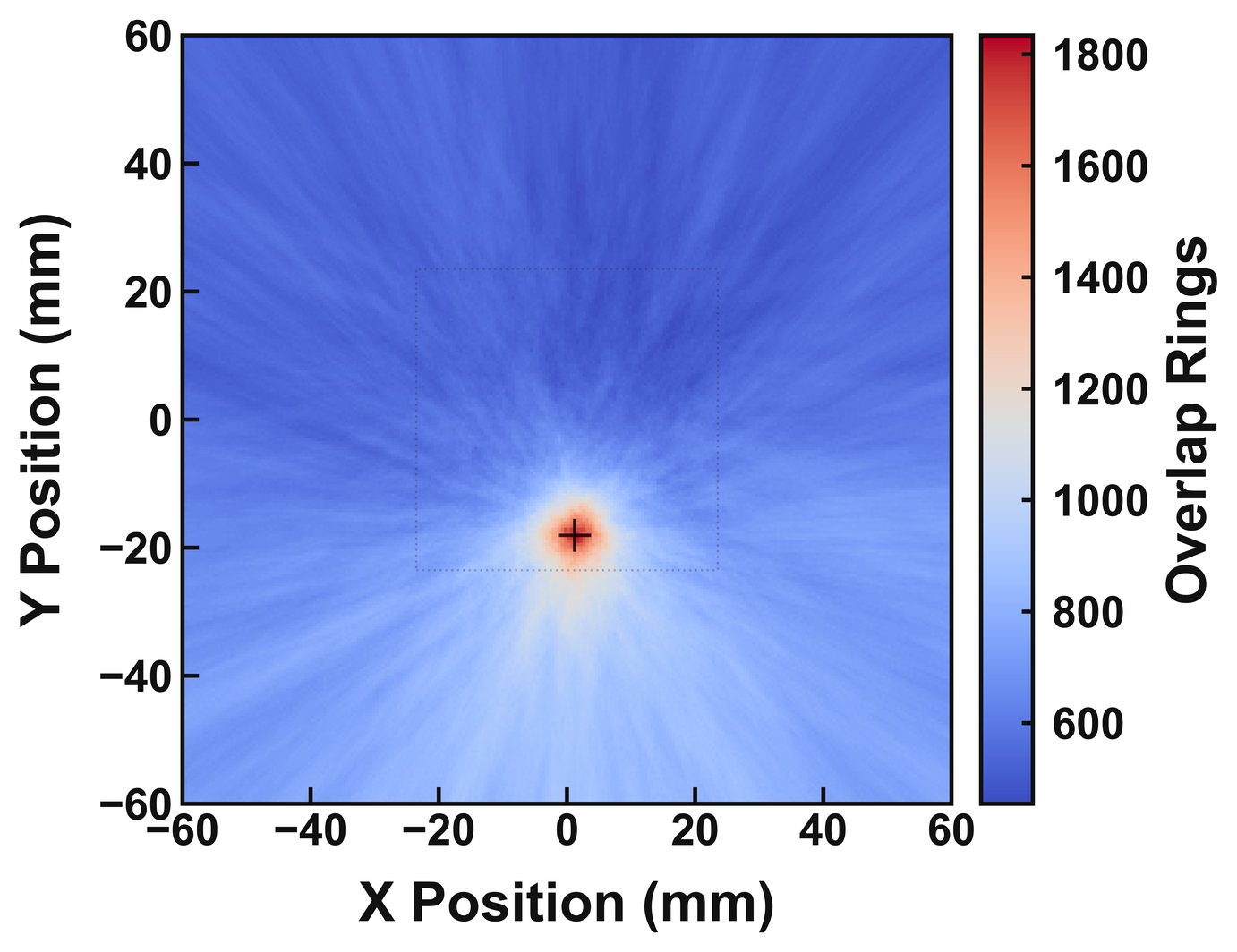}
    \caption{P$_{2}$ with SBP}
    \label{fig:SBPP2}
\end{subfigure}
\par
\begin{subfigure}[t]{0.35\columnwidth}
    \includegraphics[width=\linewidth]{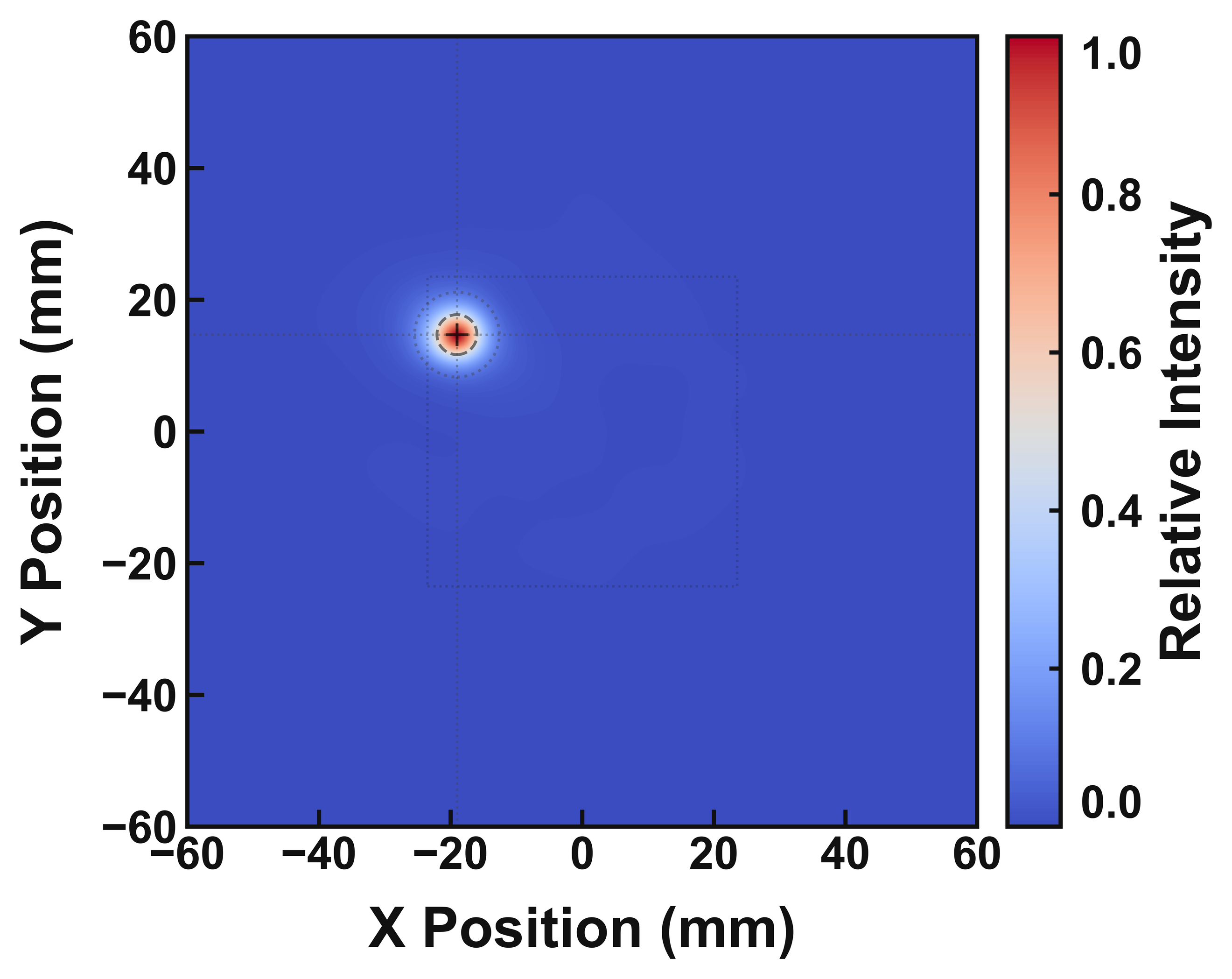}
    \caption{P$_{1}$ with MLEM}
    \label{fig:MLEMP1}
\end{subfigure}
\hspace{0.5cm}
\begin{subfigure}[t]{0.35\columnwidth}
    \includegraphics[width=\linewidth]{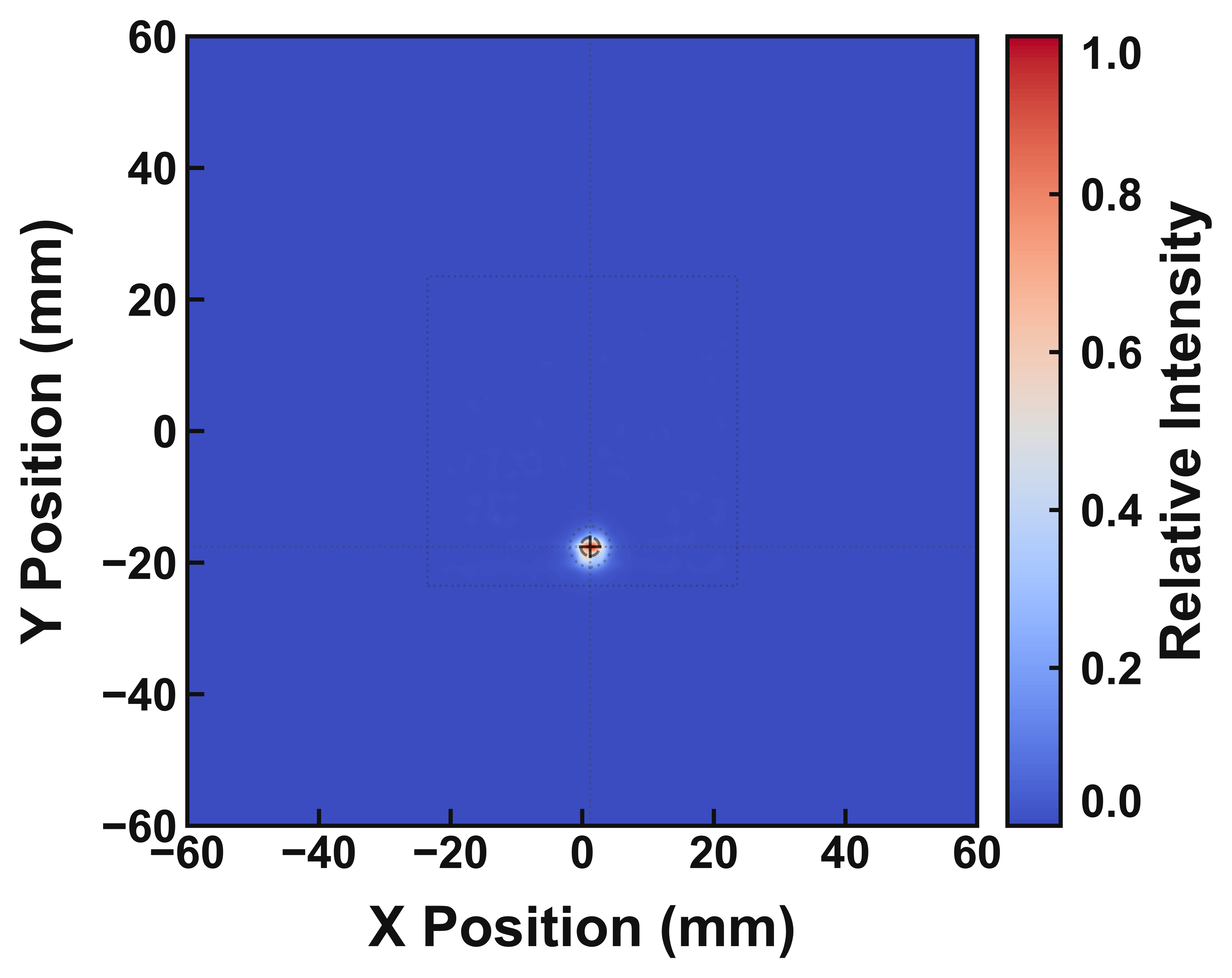}
    \caption{P$_{2}$ with MLEM}
    \label{fig:MLEMP2}
\end{subfigure}
\caption{(a): ARM distribution for 0.662 MeV events overlaid the fitting with Eq.~\ref{eq:fit_model}. (b)-(e): $^{137}$Cs source is imaged at position-1 (P$_{1}$) and position-2 (P$_{2}$) with SBP and MLEM algorithm.}
\label{fig:gamma_image}
\end{figure}

The distribution of the calculated ARM is shown in Figure~\ref{fig:arm}. It is modeled with a function consisting of a Voigt profile combined with a flat background:
\begin{equation}
    f(x) = A \cdot V(x - \mu; ~\sigma, ~\gamma) + C,
    \label{eq:fit_model}
\end{equation}
where the parameter $A$ represents the signal amplitude, and $C$ is the flat background constant. Mathematically, the Voigt profile, $V(x - \mu; ~\sigma, ~\gamma)$, is defined as the convolution of a Gaussian and a Lorentzian distribution, explicitly parameterized by the Gaussian standard deviation $\sigma$ and the Lorentzian half-width $\gamma$. Physically, the Gaussian $\sigma$ represents the angular resolution limited by the detector's intrinsic energy and spatial resolution, while the Lorentzian $\gamma$ accounts for the physical Doppler broadening of the scattering electrons in the target material.

Practically, a two-step fitting procedure is implemented in order to eliminate the severe parameter degeneracy between the slow-decaying Lorentzian tails of the Voigt profile and the flat background. First, the flat background density is estimated from the far-end tails of the ARM distribution where $|x|$>60$^{\circ}$, establishing the background constant $C$. Then, a four-parameter ($\mu$, $\sigma$, $\gamma$, and $A$) Voigt fit is executed with this background constant fixed. The fitting result is shown as the red line in Figure~\ref{fig:arm}. Minimizing the chi-squared residual of this model yields the intrinsic angular resolution, represented directly by the extracted Gaussian standard deviation $\sigma$, which is approximately 6.0$^{\circ}$. This indicates a good agreement in the angle reconstruction with the deposited energies and interaction positions in the Compton scattering mode. Furthermore, this result proves the reliability of the energy calibration, position reconstruction, and event-sequencing procedure of the developed Compton camera.

Based on the detector's evaluated response to angle and energy, the image of gamma-ray source was reconstructed. An initial estimation of the source location was obtained using a method of Simple Back-Projection (SBP) which is widely used for a rapid geometric reconstruction in the  Compton imaging technology \cite{Lee2017,Yang2026}. This method divides the imaging volume into voxels and performs the spatial reconstruction by accumulating all the voxels crossed by the Compton cones~\cite{Lee2017,Huang2021,Yang2026}. Figure~\ref{fig:SBPP1} and~\ref{fig:SBPP2} shows the source that was imaged at two different positions with the SBP algorithm where the varied source location is consistent with the realistic setup. This well validates the functionality of the Compton imaging. However, the reconstructed intensity distribution still appeared blurred, primarily due to the uncertainties of the interaction coordinates and deposited energies measured in the experiment, as well as the statistical fluctuations of the coincident events.

To suppress the projection artifacts in the SBP method and improve the imaging quality, the Maximum Likelihood Expectation Maximization (MLEM) algorithm was utilized~\cite{SheppVardi1982}. As a classic statistical iterative reconstruction method, MLEM employs the Poisson statistics to model the probability distribution of the detected coincident events. More specifically, the calibrated ARM width $\sigma$ obtained previously was used to define the angular thickness of the projected Compton cones \cite{Maxim2016MLEM,Ren2025}, thereby implementing the system uncertainties from the measurements into the probability estimation. As shown in Figure~\ref{fig:MLEMP1} and~\ref{fig:MLEMP2}, the MLEM reconstruction algorithm successfully eliminated the diffuse projection artifacts from SBP, resolving a more highlighted source area with a larger signal-to-noise ratio. The Compton source imaging with MLEM precisely aligns with the true source position. By comparing the source location change in the reconstruction with the direct measurement, we achieve a source localization accuracy of $\sim$2 mm.

\section{Environmental Tests for In-orbit Operation}
\label{sec:validation}
The Compton camera developed in this work is going to be integrated into a commercial micro-satellite for operation in space. Consider the special in-orbit environment, the detectors including the critical electronics and mechanical structure are manufactured with the aerospace-grade materials. However, following the aerospace's severe criteria, we conducted the required environmental tests at the system level prior to delivering the instrument to the satellite, including the vibration, thermal cycle and stability test~\cite{app11062659}. The details of these tests are presented in this section.

\subsection{Vibration Test}
The ground-based vibration tests are crucial because they simulate the intense mechanical stresses experienced during the launch. The dynamic forces can cause sensitive components—such as optics, electronics, wiring, and structural joints—to loosen, crack, misalignment, or even fail entirely if not properly designed and tested. We subjected the Compton camera to the controlled vibration tests on the ground which include the sine sweep and random vibration, and verify that all hardware remains intact and functional under the worst-case launch conditions. 

The instrument was tested in three orthogonal directions (X, Y and Z). Table~\ref{tab:vib} summarizes the corresponding configurations. In the regular vibration test (sine), the instrument was accelerated up to 10 g (acceleration of gravity‌, 1 g= 9.8 m/s$^{2}$) in one minute with the maximum frequency of 100 Hz. For the random vibration test, the total root mean square of the acceleration was 10.13 g and the rate was increased up to 2000 Hz.
\begin{table}[htbp]
    \centering
    \caption{Summary of the configurations for the vibration tests. \textit{f} is the vibration frequency, PSD is the power spectral distribution, RMS is the root mean square.}
    \label{tab:vib}
    \begin{tabular}{c|c|c|c|c}
    \hline
    Type & \multicolumn{4}{c}{Test Parameters}   \\
    \hline
    \multirow{4}{*}{Sine wave} & \textit{f} [Hz] & Vibration & Scan Rate & Direction \\
    \cline{2-5}
                         & 5$\sim$14 &10.14 mm & \multirow{3}{*}{4 Oct/min} & \multirow{3}{*}{X/Y/Z} \\
                         & 14$\sim$30 &8 g & &             \\
                         & 30$\sim$100 &10 g & &           \\
    \hline
    \multirow{4}{*}{Random} & \textit{f} [Hz] & PSD & RMS & Direction \\
    \cline{2-5}
                         & 20$\sim$100 &3 dB/Oct& \multirow{3}{*}{10. 13 g} & \multirow{3}{*}{X/Y/Z} \\
                         & 100$\sim$600 & 0.125 g$^{2}$/Hz & &  \\
                         & 600$\sim$2000 & -9 dB/Oct & &        \\
    \hline
    \end{tabular}
\end{table}

Figure~\ref{fig:vibration} shows the vibration evolution as the function of frequency during the tests. Upon completion of the vibration tests, we checked the instrument conditions. No deformation or damage was observed, and all the screws and connections were still tighten. In addition, we measured the muon energy spectra to monitor the detector's performance before and after the vibration tests. By comparing the MPV and sigma of MIP, the overall variation was found to be around 1\%. Therefore, we concluded that the Compton camera well passed the vibration tests.
\begin{figure}[htbp]  
\centering
    \begin{subfigure}[t]{0.32\columnwidth}
        \includegraphics[width=\linewidth]{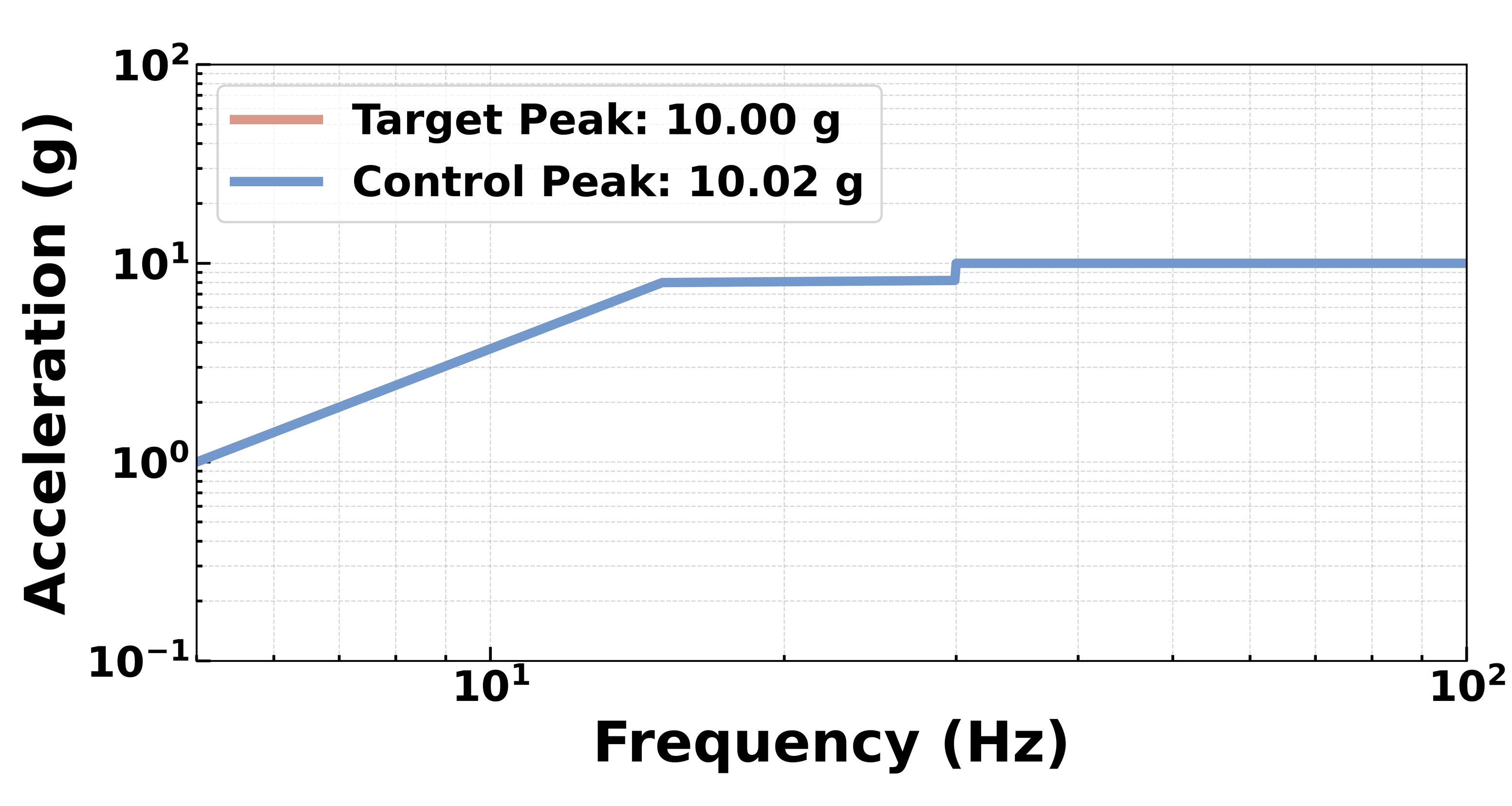}
        \caption{Sine wave in X}
    \end{subfigure}
    \hfill
    \begin{subfigure}[t]{0.32\columnwidth}
        \includegraphics[width=\linewidth]{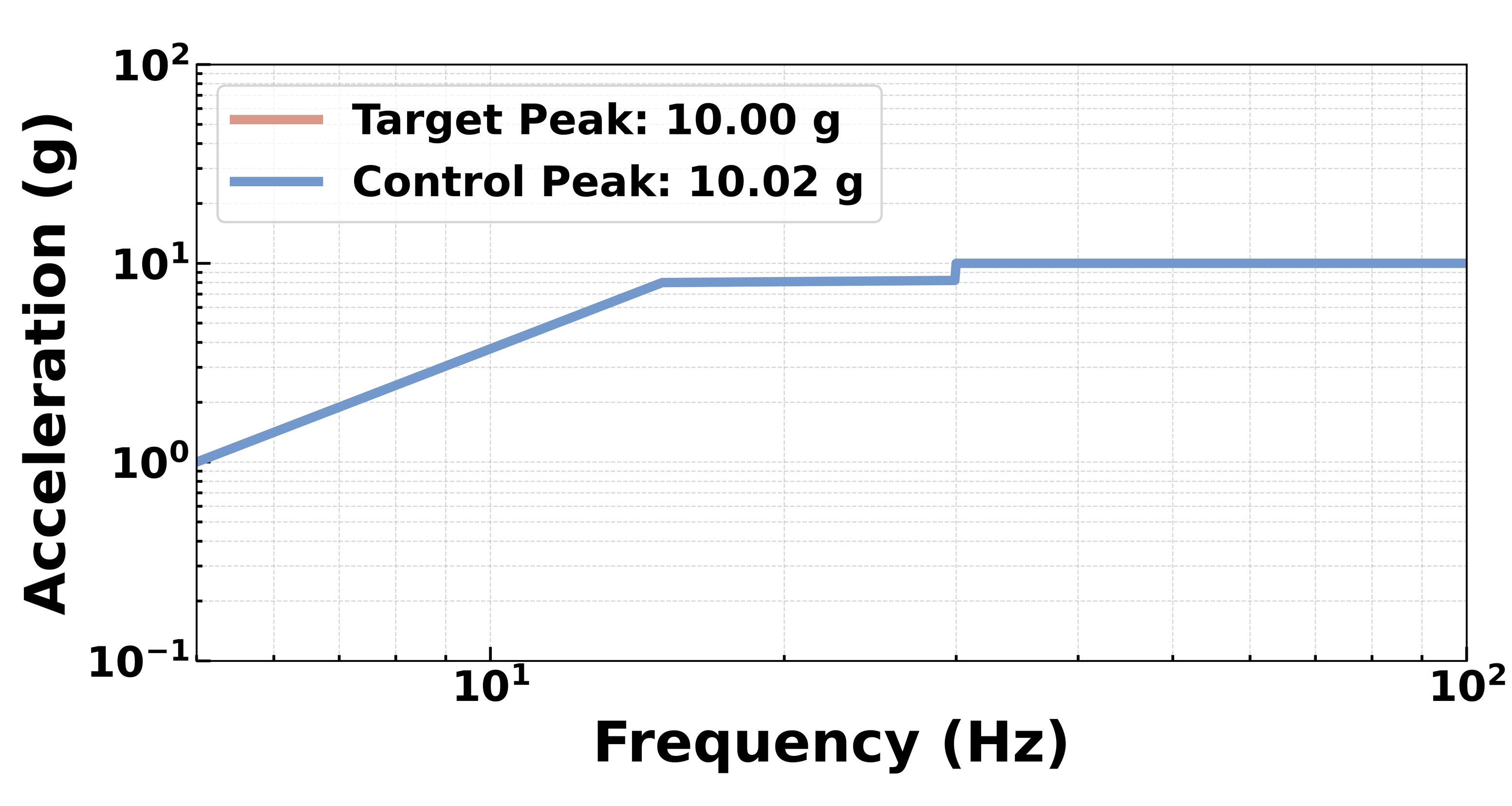}
        \caption{Sine wave in Y}
    \end{subfigure}
    \hfill
    \begin{subfigure}[t]{0.32\columnwidth}
        \includegraphics[width=\linewidth]{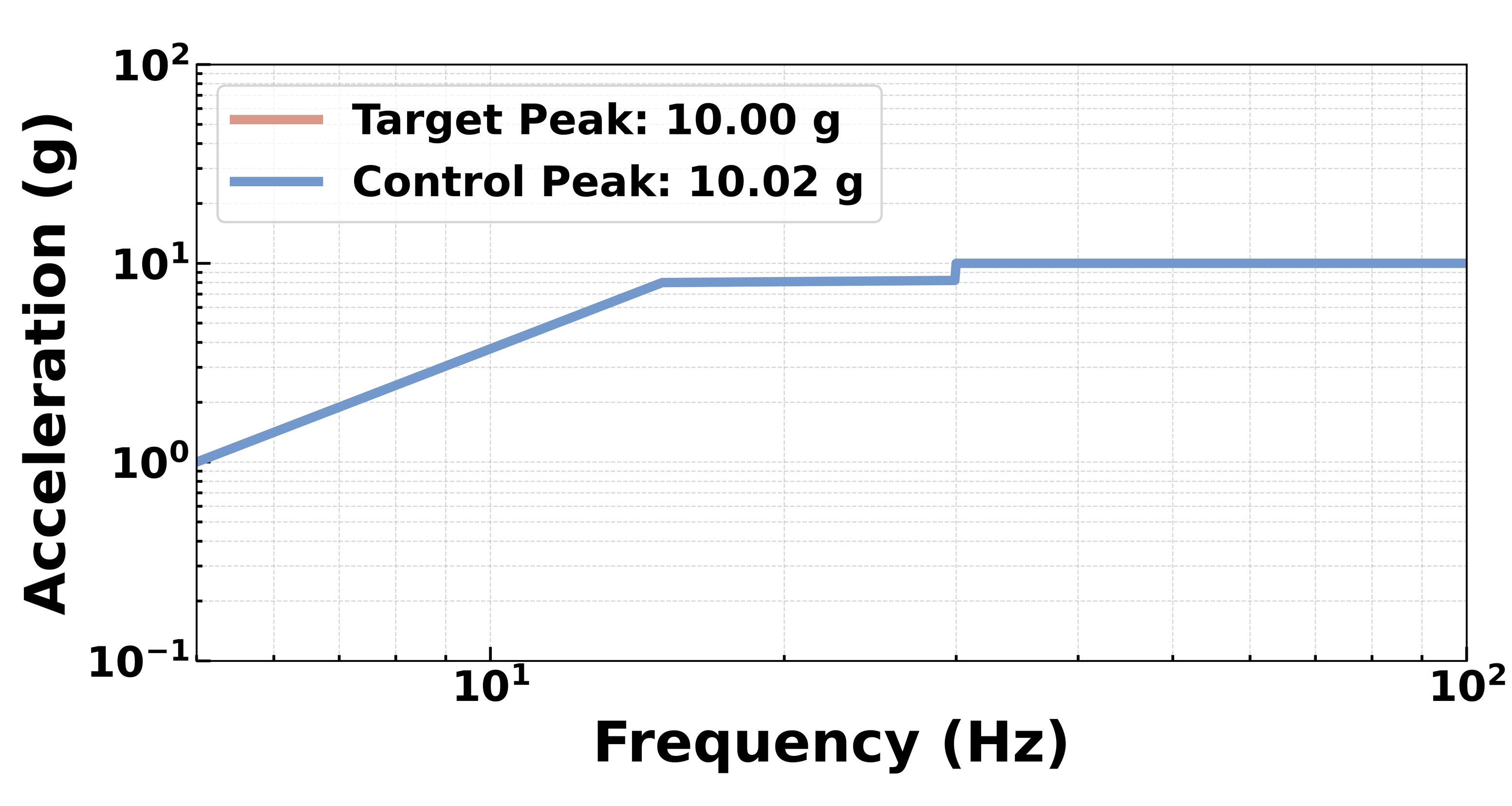}
        \caption{Sine wave in Z}
    \end{subfigure}
    \begin{subfigure}[t]{0.32\columnwidth}
        \includegraphics[width=\linewidth]{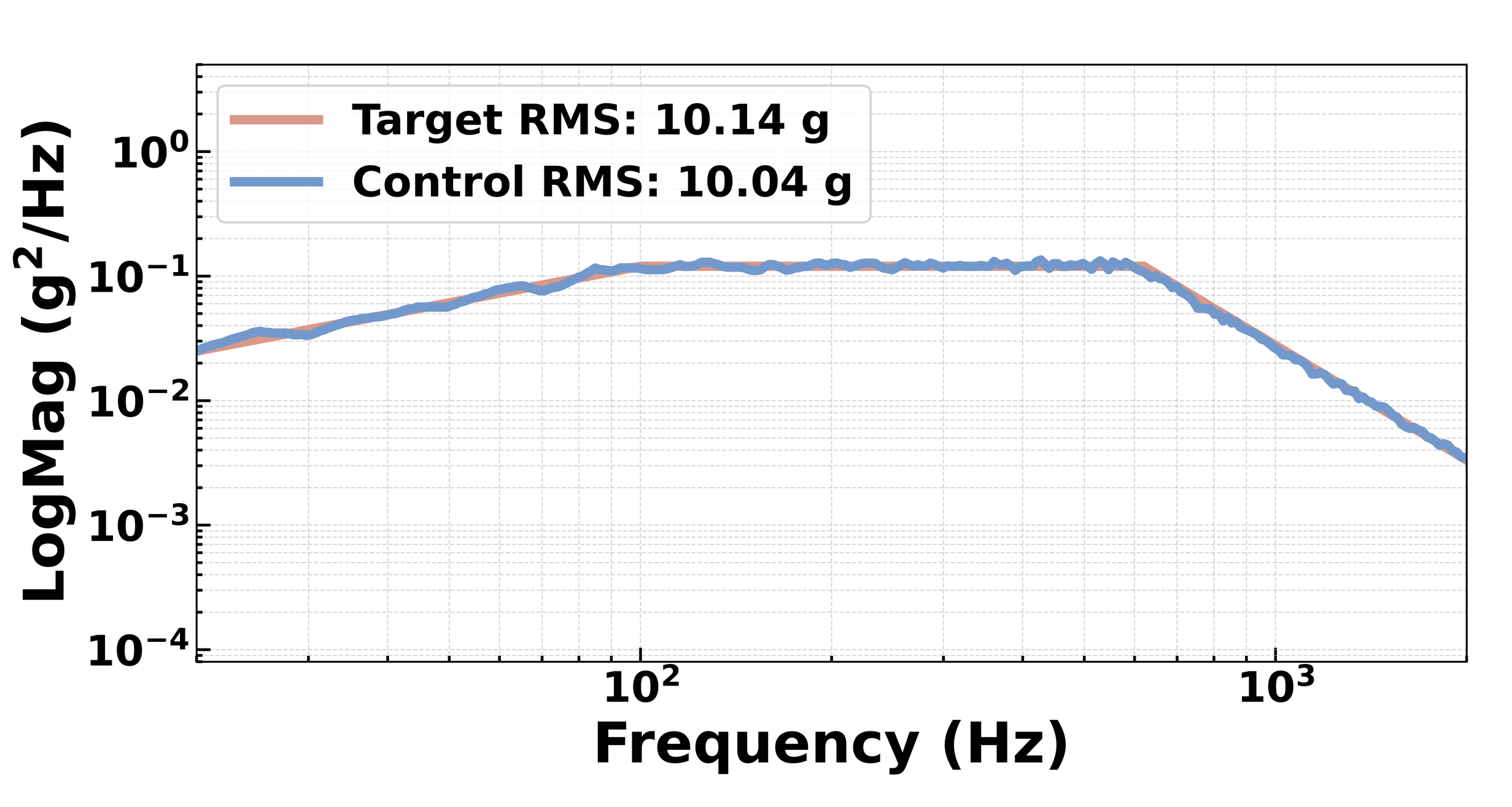}
        \caption{Random vibration in X}
    \end{subfigure}    
    \hfill
    \begin{subfigure}[t]{0.32\columnwidth}
        \includegraphics[width=\linewidth]{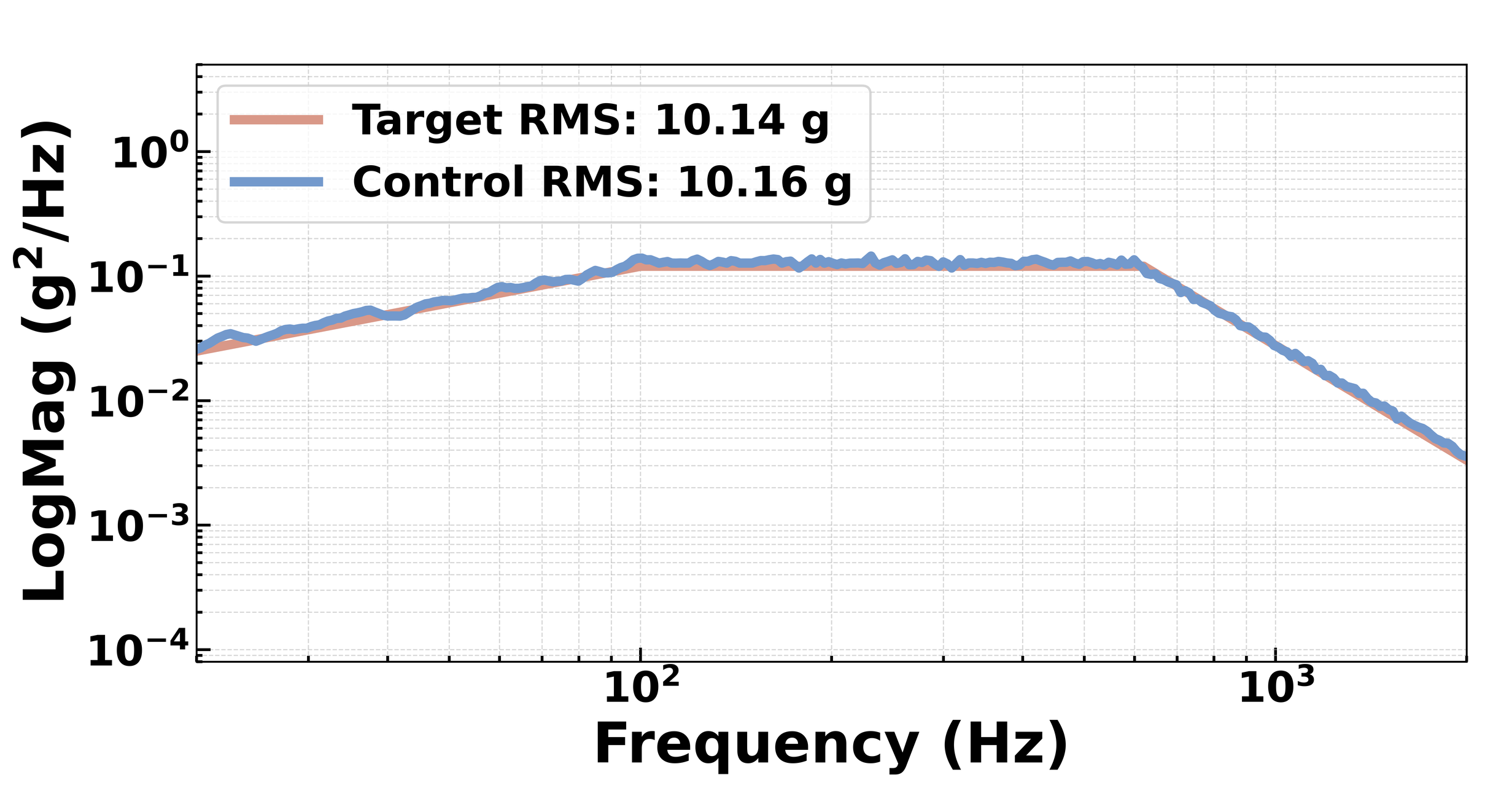}
        \caption{Random vibration in Y}
    \end{subfigure}
    \hfill
    \begin{subfigure}[t]{0.32\columnwidth}
        \includegraphics[width=\linewidth]{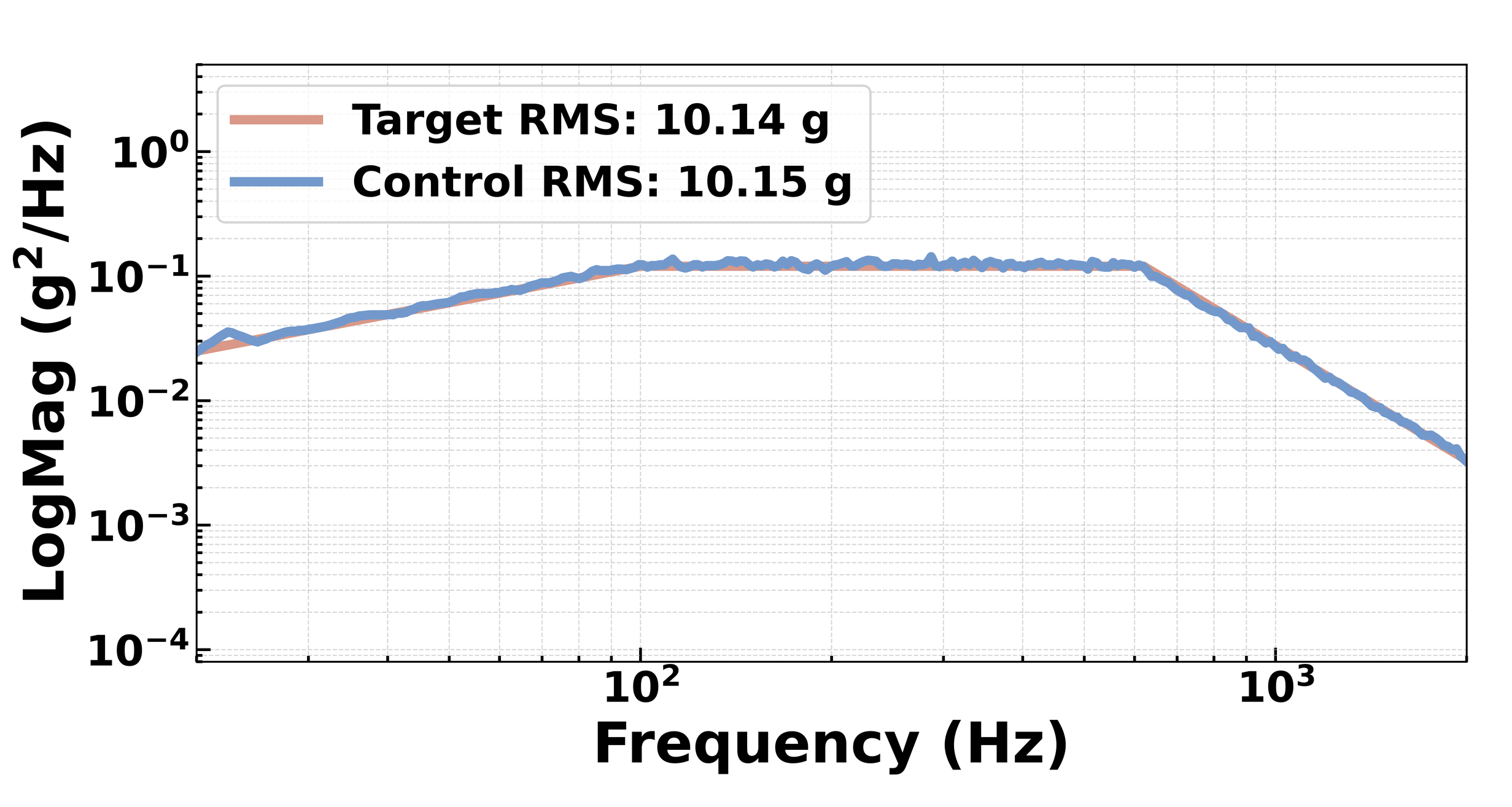}
        \caption{Random vibration in Z}
    \end{subfigure}   
\caption{Sine wave and random vibration test of the Compton camera: the red lines show the set values and blue lines are the actual readouts from the accelerometer attached.}
\label{fig:vibration}
\end{figure}

\begin{figure}[htbp]   
    \centering
    \begin{subfigure}[t]{0.8\columnwidth}
        \includegraphics[width=\linewidth]{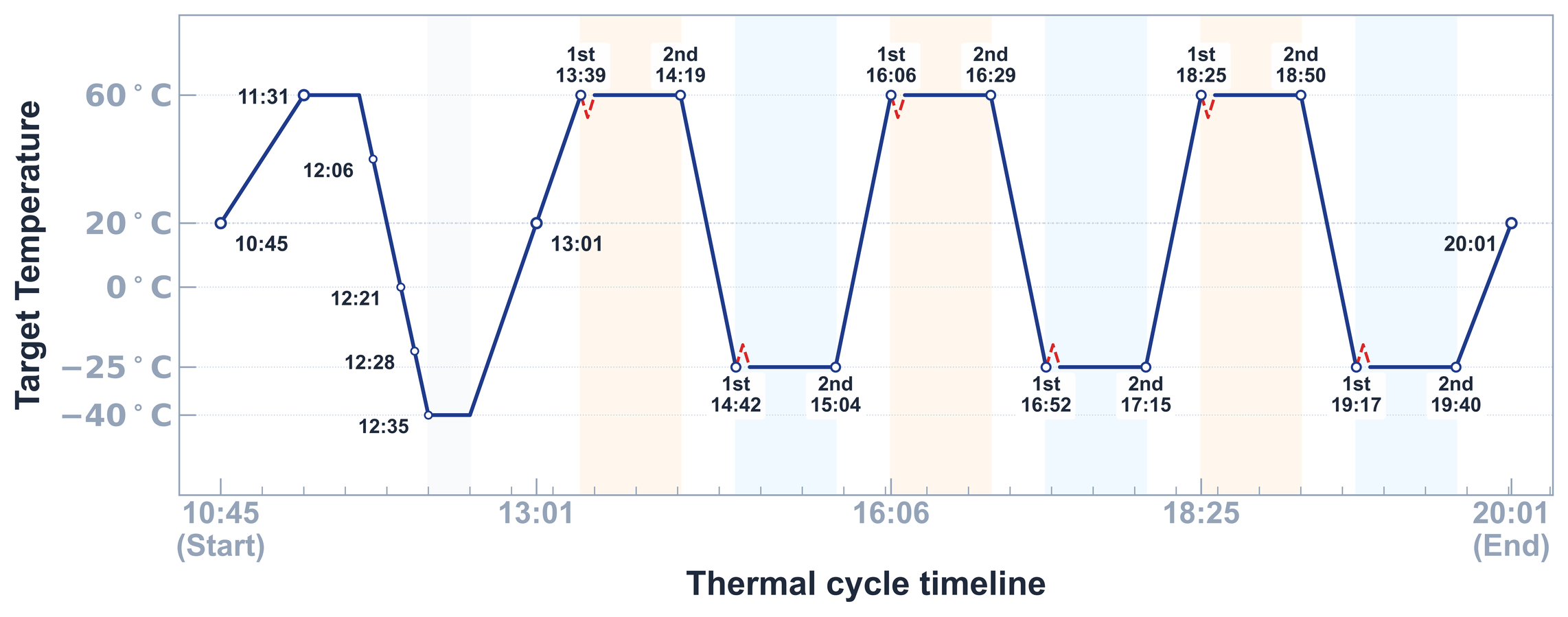}
	   \caption{Temperature evolution in one thermal cycling test.}
        \label{fig:thermal_cycles}
    \end{subfigure}
    \begin{subfigure}[t]{0.32\columnwidth}
        \includegraphics[width=\linewidth]{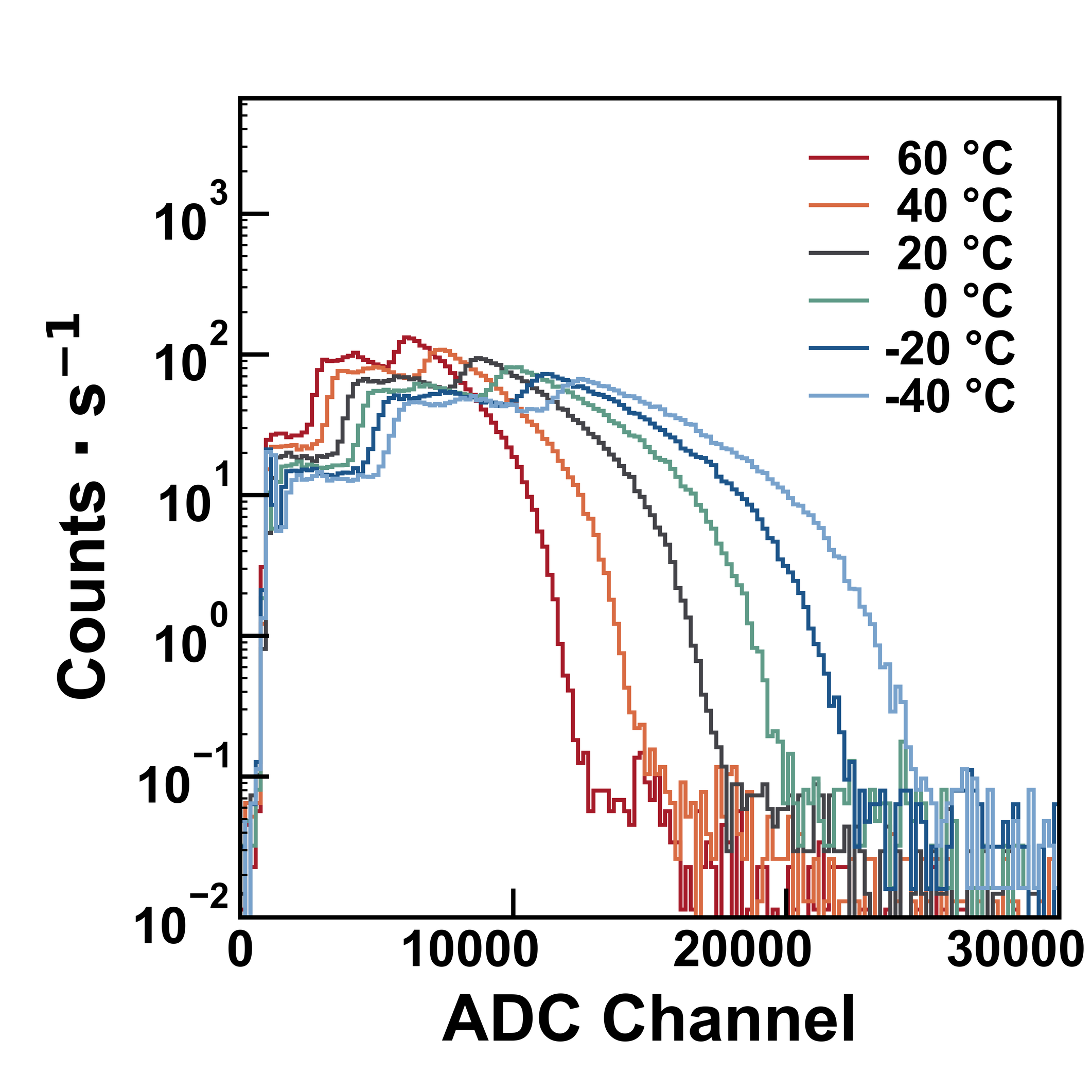}
        \caption{CH0}
        \label{fig:thermal-ch0}
    \end{subfigure}
    \hfill
    \begin{subfigure}[t]{0.32\columnwidth}
        \includegraphics[width=\linewidth]{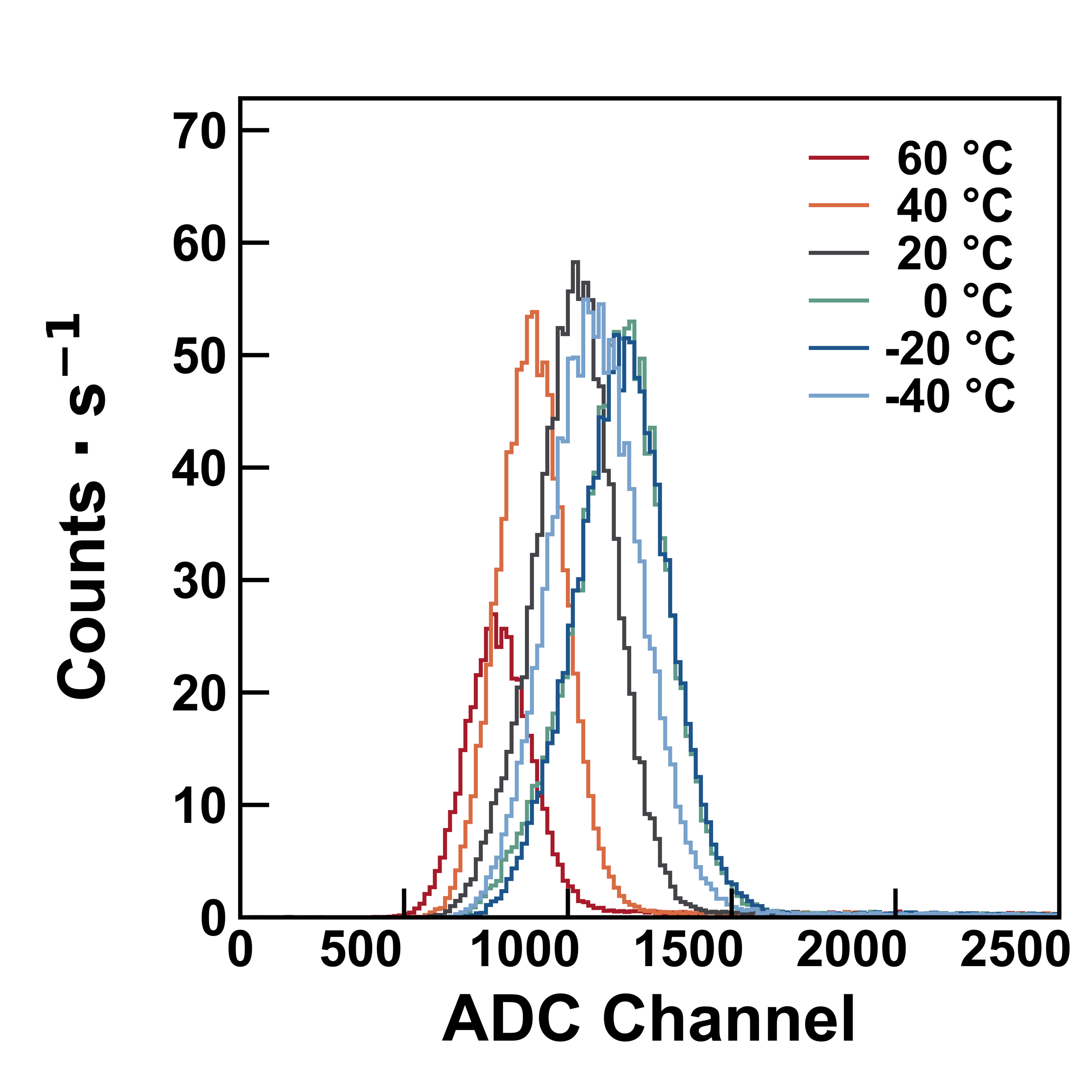}
        \caption{CH1}
        \label{fig:thermal-ch1}
    \end{subfigure}
    \hfill
    \begin{subfigure}[t]{0.32\columnwidth}
        \includegraphics[width=\linewidth]{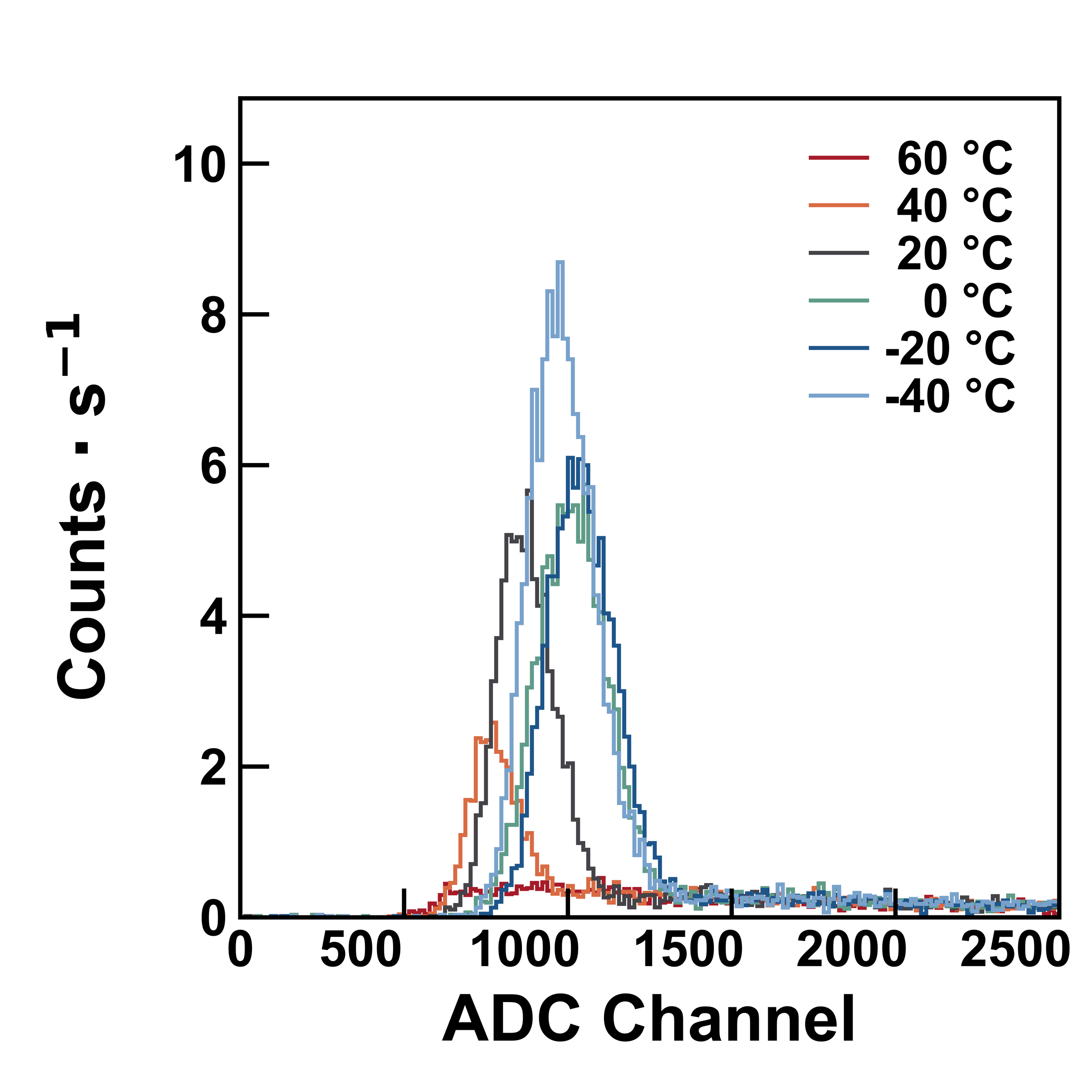}
        \caption{CH2}
        \label{fig:thermal-ch2}
    \end{subfigure}
    \caption{Thermal tests and detector's response at various temperatures.}
\end{figure}
\subsection{Thermal Test}
Thermal cycling tests are essential for satellite experiments because of temperature fluctuations that payloads may experience in orbit. In addition, due to the vacuum environment, the heat produced from the instrument can not efficiently dissipate. These effects might cause the materials to expand and contract, which can induce mechanical stress, micro-cracks, or even permanent deformation in sensitive components, solder joints, and significant electronics noise. By performing the thermal cycling tests on the ground, the instrument can be validated to withstand these harsh thermal conditions without degradation or failure, ensuring the reliability of satellite mission and the integrity of scientific data collection in space.

According to the simulation with the full payload, the temperature inside the satellite cube is expected to be in the range of -20$\sim$+40 $^{o}$C. In the ground-based thermal tests, more severe tests were conducted. Figure~\ref{fig:thermal_cycles} shows the temperature evolution in one of the serial thermal-cycle tests. The highest temperature was +60 $^{o}$C and the lowest was -40 $^{o}$C. The instrument stayed at the extreme temperatures over than half an hour in order to ensure the detector components reached the thermal equilibrium. The power supply was switched on and off to examine its functionality. 

In addition, one $^{241}$Am source was placed inside the thermal chamber together with the Compton camera, and the DAQ was run to take data in the calibration mode at various temperatures. Figure~\ref{fig:thermal-ch0} through Figure~\ref{fig:thermal-ch2} show the energy spectra of three detectors at the testing temperatures. Clearly, the energy peak of $^{241}$Am shifts along with the temperature change. It should be noted that the spectrum of CH2 at +60 $^{o}$C is not shown since the threshold of CH2 was accidentally set too high during the test which resulted in a significant cutoff on the spectrum. However, CH2 was functional at that high temperature.

The detector's response to energy varies at different temperatures due to the following facts: i) the light yield of the scintillator is temperature-dependent. LYSO and YSO material could produce different amount of photons from the same energy deposition at different temperatures. ii) the SiPM's gain is strongly correlated with the temperature. The signal amplitude from the same number of photons depends on the temperature. iii) the electronics could perform non-linearly at various temperatures. With the calibration data taken at these temperatures, we build the temperature-dependent response functions for each detector. Figure~\ref{fig:escale_tempts} shows the detector's readouts in ADC as the function of deposited energy. With Eq.~\ref{eqa:linear}, we fitted the results at each temperature, and the fitted parameters are summarized in Table~\ref{tab:para_tempt}.
\begin{figure}[htbp]   
    \centering
    \begin{subfigure}[t]{0.325\columnwidth}
        \includegraphics[width=\linewidth]{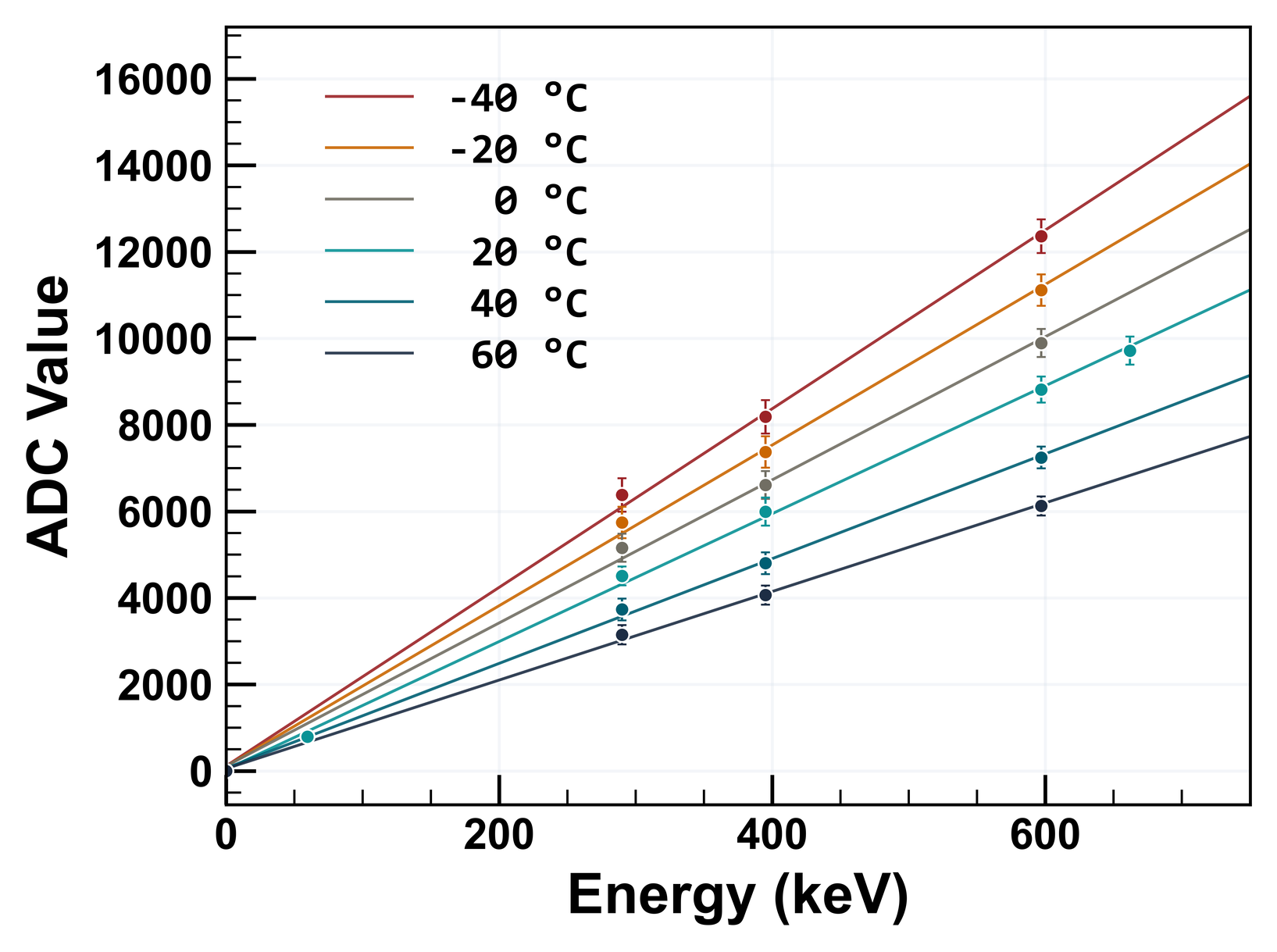}
        \caption{CH0}
    \end{subfigure}
    \hfill
    \begin{subfigure}[t]{0.325\columnwidth}
        \includegraphics[width=\linewidth]{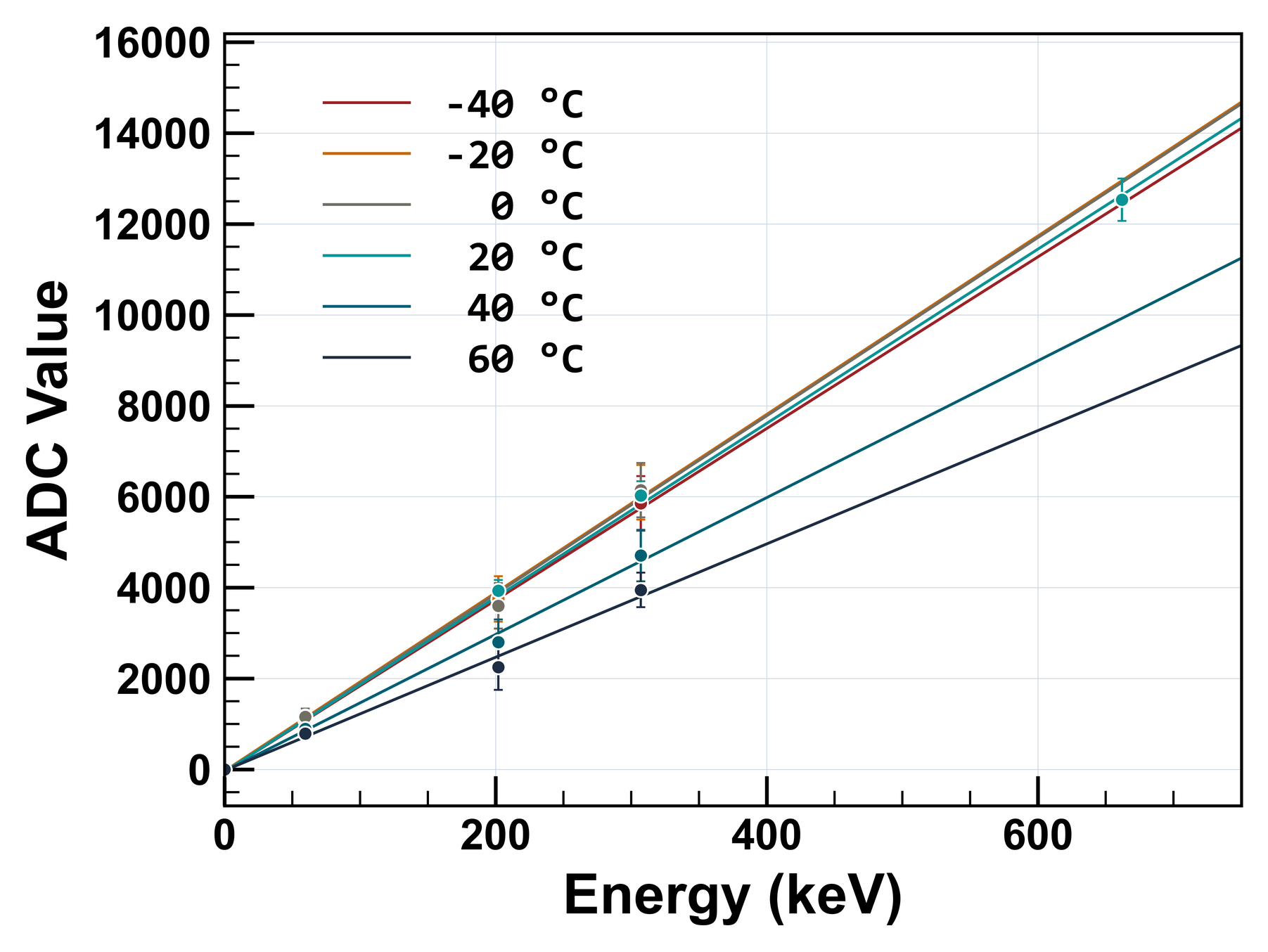}
        \caption{CH1}
    \end{subfigure}
    \hfill
    \begin{subfigure}[t]{0.325\columnwidth}
        \includegraphics[width=\linewidth]{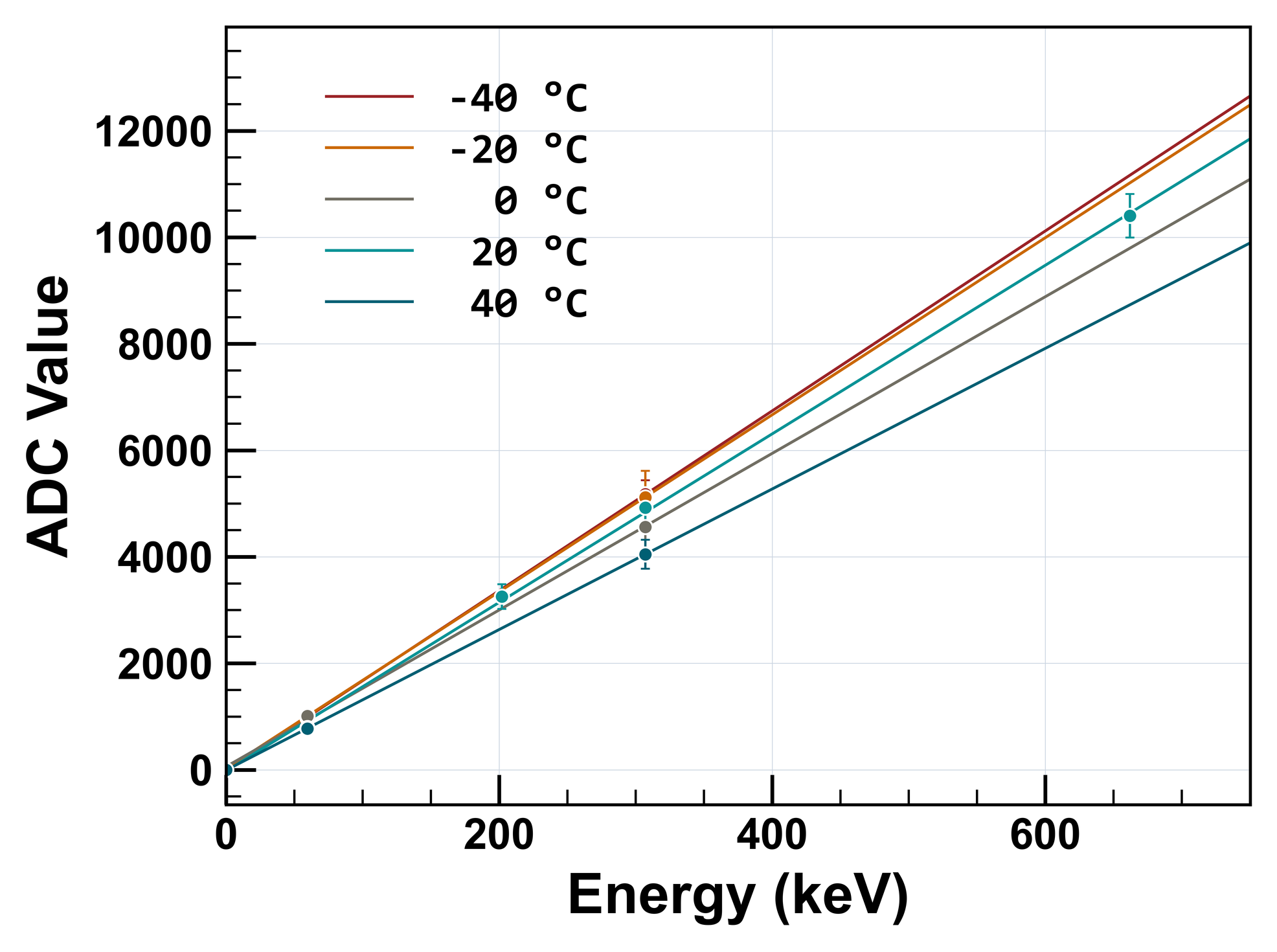}
        \caption{CH2}
    \end{subfigure}
    \caption{Temperature-dependent response to energies for each detector.}
    \label{fig:escale_tempts}
\end{figure}
\begin{table}[htbp]
    \centering
    \caption{Summary of the fitted parameters to build each detector's response to energy at various temperatures.}
    \label{tab:para_tempt}
    \begin{tabular}{c|c|c|c|c|c|c}
    \hline
    \multirow{2}{*}{Temperature} & \multicolumn{2}{c|}{CH0} & \multicolumn{2}{c|}{CH1} & \multicolumn{2}{c}{CH2}  \\
    \cline{2-7}
         & a & b & a & b & a & b    \\
     \hline
     -40 $^{o}$C & 20.6 & 114.9 & 18.9 & -47.3 & 16.9 & -14.1 \\
     -20 $^{o}$C & 18.6 & 107.0 & 19.6 & -34.6 & 16.6 & 19.9  \\
       0 $^{o}$C & 16.5 & 114.8 & 19.6 & -61.2 & 14.7 & 60.6  \\
      20 $^{o}$C & 14.8 & 36.4 & 19.2 & -49.4 & 15.8 & -20.6  \\
      40 $^{o}$C & 12.1 & 65.9 & 15.0 & -39.2 & 13.2 & -4.9  \\
      60 $^{o}$C & 10.2 & 53.7 & 12.5 & -23.6 & n.a. & n.a.   \\
    \hline
    \end{tabular}
\end{table}

\subsection{Stability Test}
The Compton camera is supposed to work in orbit for years, then the stability test is essential. With the coincidence trigger for muons, we continuously operated the detector over two weeks when the environmental temperature was controlled by an air conditioner, with a variation of around $\pm$1 $^{\circ}$C. Six independent muon data sets were collected. From each data set, the MIP energy spectrum is reconstructed. Using the Landau fitting discussed in Section~\ref{sec:performance}, the MPV of MIPs as well as the corresponding spread were monitored from run to run.

Figure~\ref{fig:mip_time} shows the evolution of MPV fitted on the spectrum from the raw data, spectrum after the spatial non-uniformity and incident-angle correction. The overlaid error bars are the MPV spread ($\pm$1$\sigma$). The conclusion is the Compton camera performed reliably stable under long-term continuous operation. The overall fluctuation is $\sim$1\% which makes the future's in-orbit operation feasible and reliable.
\begin{figure}[htbp]  
    \centering
    \begin{subfigure}[t]{0.42\columnwidth}
        \includegraphics[width=\linewidth]{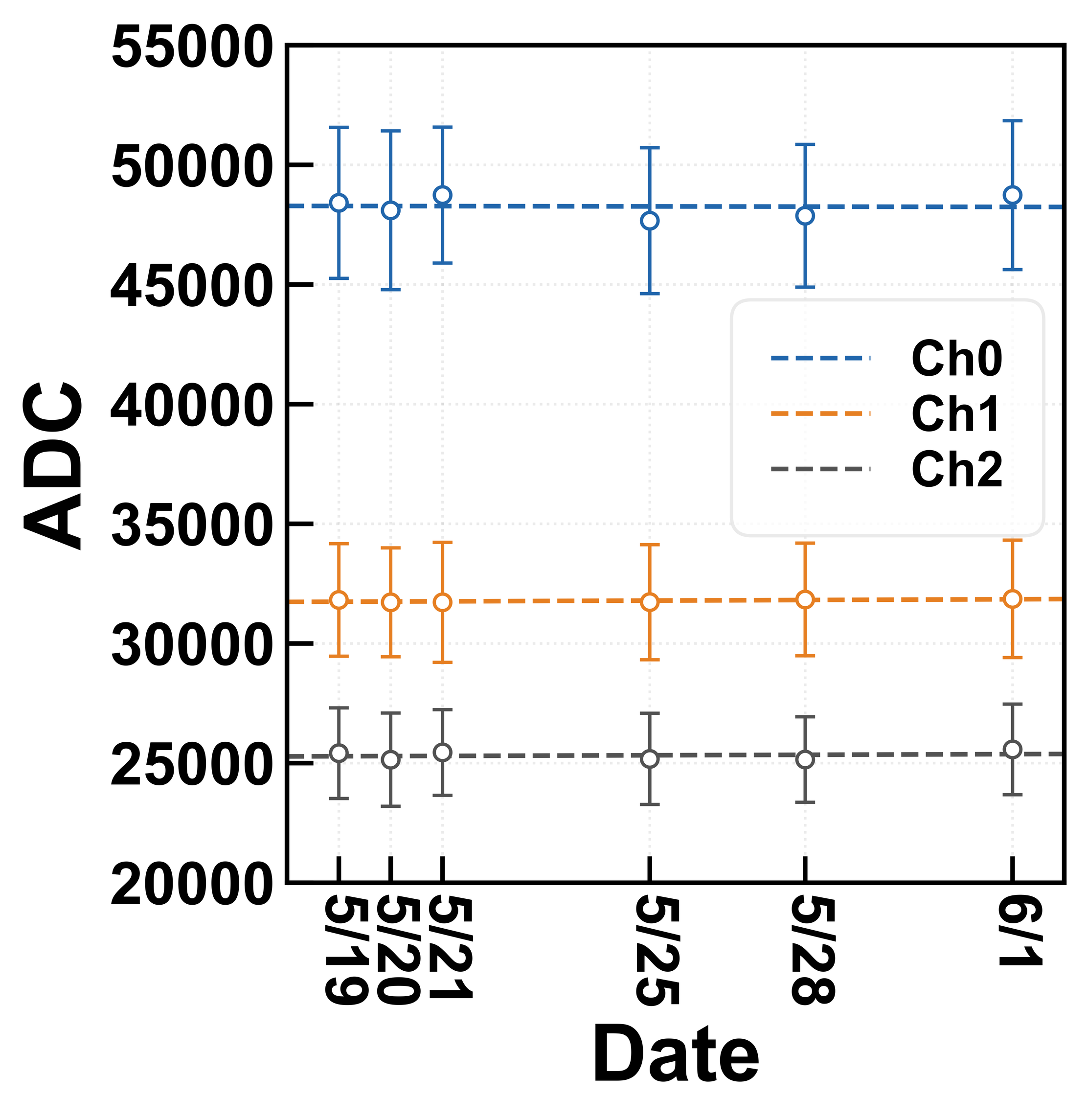}
        \caption{}
    \end{subfigure}
    \hspace{1.0cm}
    \begin{subfigure}[t]{0.42\columnwidth}
        \includegraphics[width=\linewidth]{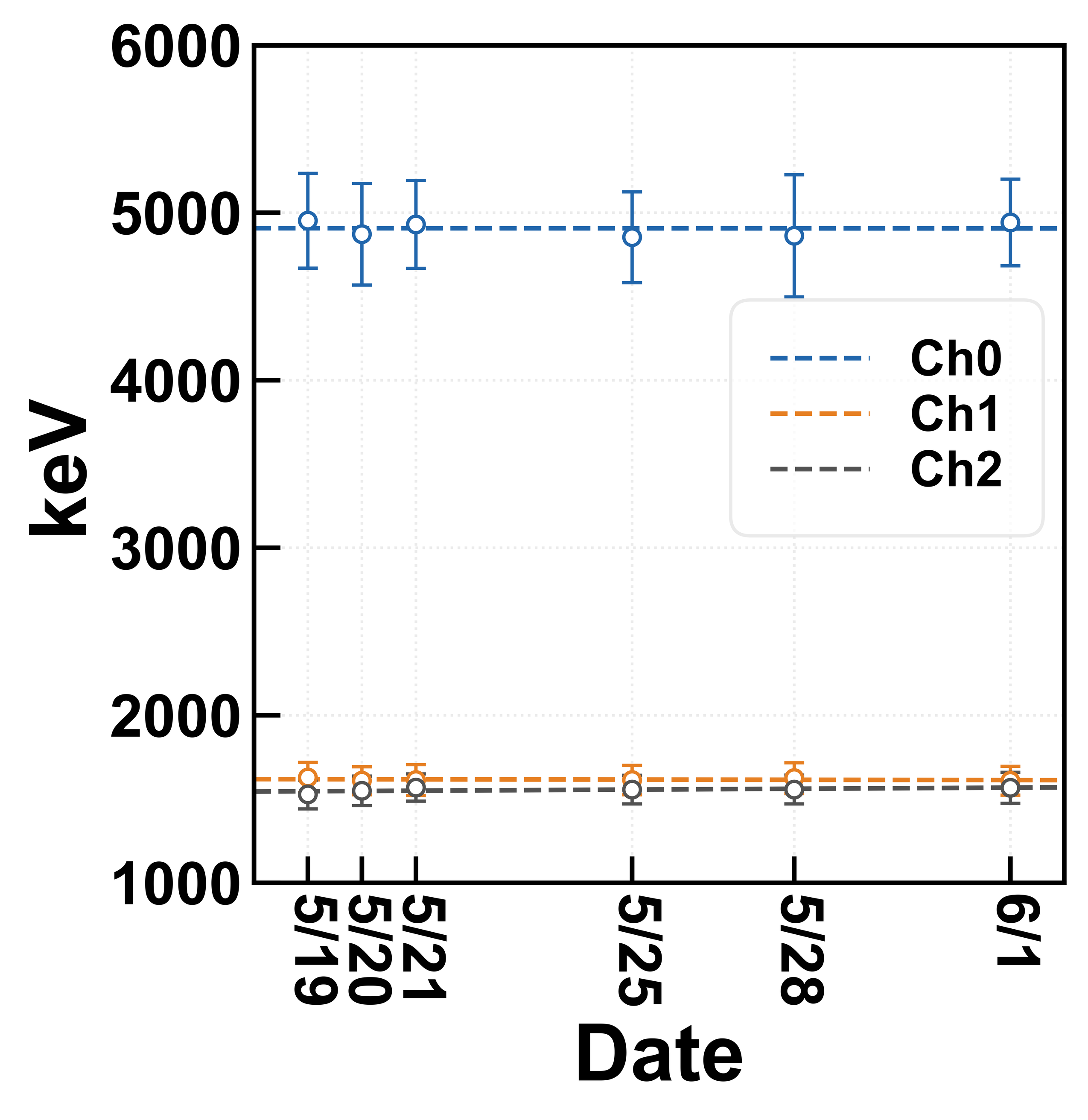}
        \caption{}
    \end{subfigure}
    \caption{Muon MPV and standard deviation evolution in the stability test: (a) from the raw ADC spectra; (b) from the spectra after the spatial and incident-angle correction.}
    \label{fig:mip_time}
\end{figure}

\section{Conclusions}
\label{sec:conclusion}
This paper presents the development of a novel Compton camera using three YSO/LYSO scintillator layers, with a focus on the calibration and validation for the detection of MeV-energy gamma-rays in space.
With the dedicated calibrations, the Compton camera performs with an energy resolution of $\sim$5.5\% for MIPs and better than 4\% for 0.662 MeV gamma-rays, together with an extremely good energy linearity in the energy range up to a few MeV. Building on this novel Compton camera, the reconstruction algorithms have been developed. Validated by the cosmic muons and radioactive gamma source, the Compton camera achieves the position resolution of $\sim$1 mm and demonstrates its good ability of gamma-ray source reconstruction with the technology of Compton imaging. Furthermore, the critical environmental tests for the Compton camera operated in space have been conducted and well passed. All these performances are validated to meet the design requirements.

The developed Compton camera has been delivered to piggyback a commercial satellite which is planned to be launched by the year of 2026. The Compton camera will be maintained to mainly point to the Galactic center during the operation in space, and its in-orbit performance as well as the further optimization of reconstruction algorithm for the gamma-ray observation will be reported separately afterwards.

\acknowledgments

This work is supported in part by National Key R\&D Program of China (No. 2025YFF0521800), Double Top-class grant from Shanghai Jiao Tong University and Shanghai Science and Technology Innovation Program (22DZ2229008). We are thankful for the support from the State Key Laboratory of Dark Matter Physics and Yangyang Development Fund. We thank Jinglai Liu and Haiguang Xu for the help to initialize the project; Wei Hu and Chunlei Qu for the helpful discussions on the electronics; Yanfeng Jing for the effective coordination on the payload integration with the satellite.

\bibliographystyle{JHEP}
\bibliography{biblio.bib}

\end{document}